\documentclass[useAMS,usenatbib]{mn2e}
\usepackage{amsmath}
\usepackage{amssymb}

\usepackage[pdftex]{graphicx}
\usepackage{epstopdf}
\usepackage[english]{babel}
\usepackage{subfigure}
\usepackage{array,multirow}
\usepackage{underscore}
\usepackage{natbib}
\usepackage{float}

\usepackage{array}
\newcolumntype{L}[1]{>{\raggedright\let\newline\\\arraybackslash\hspace{0pt}}m{#1}}
\newcolumntype{C}[1]{>{\centering\let\newline\\\arraybackslash\hspace{0pt}}m{#1}}
\newcolumntype{R}[1]{>{\raggedleft\let\newline\\\arraybackslash\hspace{0pt}}m{#1}}

\def\mnras{MNRAS}
\def\apj{ApJ}
\def\aj{AJ}
\def\aap{A\&A}
\def\aaps{A\&AS}
\def\apjl{ApJL}
\def\apjs{ApJS}

\def\pasp{PASP}

\title[Little size evolution in MACSJ0717]{The most massive galaxies in clusters are already fully grown at $z \sim 0.5$ }
\author[L. J. Oldham, R.C.W. Houghton, Roger L. Davies]{L. J. Oldham$^{1}$\thanks{E-mail: loldham@ast.cam.ac.uk}, R.C.W. Houghton$^{2}$, Roger L. Davies$^{2}$
\\
$^{1}$ Institute of Astronomy, University of Cambridge, Madingley Road, Cambridge CB3 0HA, UK \\
$^{^2}$ Physics Department, University of Oxford, Denys Wilkinson Building, Keble Road, Oxford OX1 3RH, UK}

\pubyear{2015}

\begin{document}
\label{firstpage}
\maketitle

\setcounter{page}{1}
\begin{abstract}

By constructing scaling relations for galaxies in the massive cluster MACSJ0717.5 at $z=0.545$ and comparing with those of Coma, we model the luminosity evolution of the stellar populations and the structural evolution of the galaxies. We calculate magnitudes, surface brightnesses and effective radii using HST/ACS images and velocity dispersions using Gemini/GMOS spectra, and present a catalogue of our measurements for 17 galaxies. We also generate photometric catalogues for $\sim 3000$ galaxies from the HST imaging. With these, we construct the colour-magnitude relation, the fundamental plane, the mass-to-light versus mass relation, the mass-size relation and the mass-velocity dispersion relation for both clusters. We present a new, coherent way of modelling these scaling relations simultaneously using a simple physical model in order to infer the evolution in luminosity, size and velocity dispersion as a function of redshift, and show that the data can be fully accounted for with this model. We find that (a) the evolution in size and velocity dispersion undergone by these galaxies between $z \sim 0.5$ and $z \sim 0$ is mild, with $R_e(z) \sim (1+z)^{-0.40\pm0.32}$ and $\sigma(z) \sim (1+z)^{0.09 \pm 0.27}$, and (b) the stellar populations are old, $\sim 10$ Gyr, with a $\sim 3$ Gyr dispersion in age, and are consistent with evolving purely passively since $z \sim 0.5$ with $\Delta \log M/L_B = -0.55_{-0.07}^{+0.15} z$. The implication is that these galaxies formed their stars early and subsequently grew dissipationlessly so as to have their mass already in place by $z \sim 0.5$, and suggests a dominant role for dry mergers, which may have accelerated the growth in these high-density cluster environments.

\end{abstract}
\begin{keywords}
galaxies: elliptical and lenticular, CD - galaxies: evolution - galaxies: clusters: individual: MACSJ0717.5 - galaxies: kinematics and dynamics
\end{keywords}

\section{Introduction}
\label{sec:introduction}

Observations of ETGs at high redshifts indicate that they are massive, bright and compact \citep[e.g.][]{vanDokkum2008}, and must therefore evolve through a series of minor mergers and accretion events into the large, passive systems that we see today. However, they also obey tight scaling relations, such as the fundamental plane (FP), out to redshifts $z \sim 1$ \citep[e.g.][]{holden10}, with a small scatter that is much more in keeping with stellar populations that formed early, then faded passively. As such, there remain a number of unanswered questions about how these systems have evolved: for instance, when did they undergo most of their evolution? Do all ETGs experience significant size growth? Luckily, ETGs retain observable imprints of their past which help us to answer these questions.

The FP \citep{djorgovski,dressler87} is one such tool. The virial theorem predicts the existence of a plane of the form 
\begin{equation}
 \log R_e = \alpha \log \sigma + \beta \langle\mu_e\rangle + \gamma,
\end{equation}
where $R_e, ~\sigma$ and $\langle\mu_e\rangle$ are the half-light radius, velocity dispersion and effective surface brightness (the mean surface brightness within $R_e$) respectively, with $\alpha = 2$ and $\beta = 1$, which should be tightly obeyed by all galaxies. The fact that ETGs fall on such a plane implies a high degree of uniformity in how these systems have evolved, but the tilt of the FP relative to the virial prediction -- with $\alpha = 1.24$ and $\beta = 0.33$ in the local Universe \citep{jorgensen95} -- also indicates some systematic variation across the plane, which may be either structural -- with, for instance, some mass dependence in the luminous-to-dark matter ratio or the mass profile shape -- or related to the stellar populations, with more massive galaxies having more bottom-heavy inital mass functions (IMFs). The evolution of the FP therefore holds a great deal of information about the evolution of the structure and stellar populations of ETGs themselves.

In the past, the evolution of the zeropoint of the FP has been taken as evidence for the passive fading of the stellar populations and used to measure stellar age \citep[e.g.][]{vandokkum96,bender98,vanDokkum2003,jorgensen06}, generally implying a mean star formation redshift $z_f > 2$. However, the FP is also sensitive to evolution in velocity dispersion and size, and while more recent studies have attempted to account for this, no consensus has yet been reached on the strength of this evolution. For instance, \citet{saglia10} measured the structural evolution of a sample of field galaxies out to $z \sim 0.9$ and found it to be significant, with its inclusion in the zeropoint analysis increasing the stellar age by 1-4 Gyr (depending on morphology and redshift). On the other hand, \citet{JorgensenChiboucas2013} and \citet{Jorgensen2014} found only very small differences in size and velocity dispersion for cluster galaxies across similar redshifts -- roughly one-third that of \citet{saglia10} -- and made the suggestion that structural evolution may depend on environment, with accelerated growth in dense clusters. However, \citet{Newman2014} found no evidence for size differences between field and cluster galaxies at $z = 1.8$ -- indicating that any changes must imprint themselves in a narrow redshift window -- while \citet{Valentinuzzi2010} identified large numbers of compact galaxies in clusters at $0 < z < 1$, which suggests the evolutionary scheme of ETGs in clusters is diverse. The question of how and when any growth occurs, then, and how it relates to the stellar populations, remains open, and it is important to try to answer this further using independent samples and methods.

Another extremely simple but useful scaling relation that can be used to give a measure of the ETG formation epoch is the colour--magnitude relation \citep[CMR; ][]{baum, sandage,visvanathan}, in which the ETGs fall along a tight red sequence, whose intrinsic scatter is mainly determined by the distribution of stellar age. In contrast to the FP, which contains information on dark matter content and galaxy structure, the CMR depends almost wholly on the properties of the stellar populations. Several studies out to redshifts $z \sim 1.3$ have shown the intrinsic scatter about the CMR to be consistently small \citep[generally $< 0.1$ mag: see][]{stanford95, ellis97,bower98,stanford98,mei}, suggestive of generally old stellar populations, with a redshift of formation $z_f > 2$ and a small spread in age. However, it is hard to disentangle these two degenerate factors, given that the scatter decreases with both increasing age -- as stars become asymptotically redder -- and increasing synchronicity -- as stars with similar ages have similar colours. It is therefore possible for a stellar population with recent, synchronised star formation to have the same small scatter as one in which the star formation happened longer ago but was dispersed. Clearly, if this degeneracy can be broken, it can provide informative complementary constraints on the FP.

The aim of this work is to construct these scaling relations for the cluster MACSJ0717.5+3745 (hereafter MACSJ0717) at z=0.545, and to compare with those of the Coma galaxies to investigate their evolution in terms of galaxy structure and stellar populations. The paper is organised as follows: in Section \ref{sec:data} we introduce the data and explain our reduction methods; in Section \ref{sec:results} we construct the scaling relations and in Section~\ref{sec:SPS} we interpret these in terms of stellar population models. Section \ref{sec:discussion} and Section \ref{sec:conclusions} then give a discussion and summary. We assume $\Omega _m = 0.3, \Omega _{\Lambda} = 0.7$ and $H_0 = 71 \text{kms}^{-1}\text{Mpc}^{-1}$ throughout, and calculate magnitudes in the AB system.

\section{Data sources and reduction}
\label{sec:data}
\subsection{Sources}

For both MACSJ0717 and Coma, we construct the colour-magnitude relation (CMR), the FP and mass-to-light versus mass (MLM) relation and the mass-size and mass-velocity dispersion relations. We therefore require colours, surface photometry and kinematics in each case.

For MACSJ0717, we measure photometry using archival HST/ACS images that were observed as part of the Cluster Lensing And Supernova survey with Hubble \citep[CLASH;][]{Postman2012}. For size and surface brightness measurements, we use the F475W (exposure time: 4064 s) and F625W bands (exposure time: 4128 s) in order to bracket the Balmer break in the rest frame. For catalogue generation using SExtractor \citep{Bertin1996}, we additionally use the multiband image (summing all 16 CLASH filters)  as the reference image for object detection (as explained in more detail in Section 2.2). These data are available from the CLASH archive, and have been previously corrected for galactic extinction and redrizzled to a pixel scale of $0.065''$/pixel. We measure velocity dispersions using Gemini/GMOS spectra, available in the Gemini archive for a subsample of 31 galaxies. These spectra, along with associated flat and bias frames, were taken over four dates between 04/02/2003 and 02/03/2003, using GMOS in multi-object mode ith the B600_G5303 grating (which has a resolution $R = 1688$ for a 0.5$''$ slit width at 461nm), as part of the science program GN--2002B--Q--44.

For Coma, we use the kinematic data from \cite{jorgensen99}; integrated photometry, observed in the Johnson $U$ and $V$ bands, from \cite{terlevich01}, and the surface photometry in the Gunn $r'$-band from \cite{jorgensen95}. We refer the reader to these papers for further information, though we note that the photometry has been previously corrected for galaxtic extinction, and that we convert the Gunn $r'$ photometry to AB magnitudes using the corrections listed in \citet{FreiGunn1994}.

\subsection{Photometry}

\begin{figure*}
 \centering
\includegraphics[trim=80 380 110 70,clip,width=0.9\textwidth]{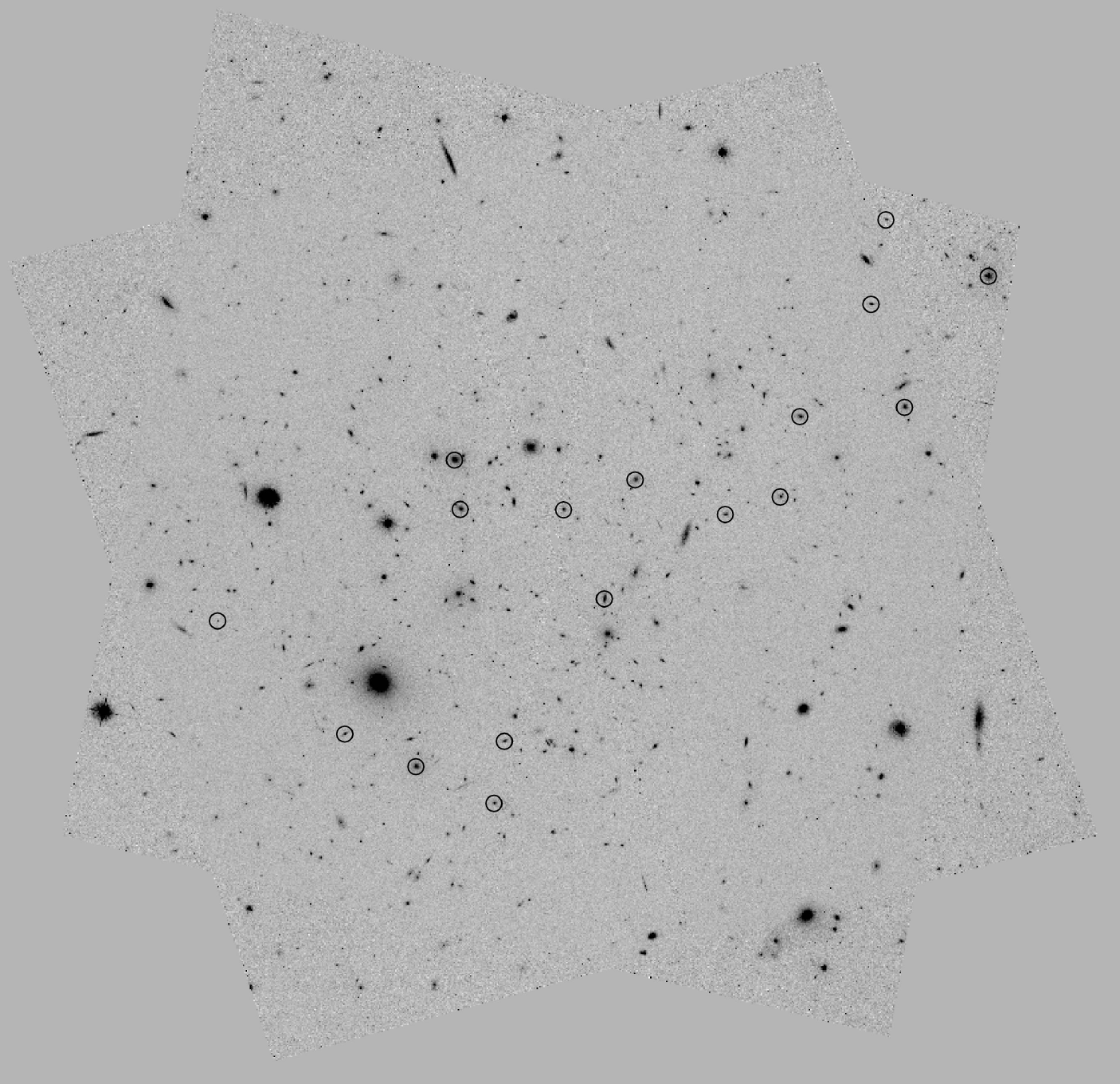}
\caption{HST/ACS F625W image of MACSJ0717, with a pixel scale of 0.065$''$/pixel and dimensions 325$\times$325$''$; black circles show the final sample of 17 galaxies.}
\label{fig:hst}
\end{figure*}

The CLASH database  \citep{Postman2012} provides a SExtractor-generated catalogue of isophotal magnitudes for $\sim 8000$ objects detected in the HST images; however, in order to calculate colours precisely for the CMR and avoid under-estimating the flux of the largest low-surface-brightness galaxies, we use SExtractor to generate a catalogue of integrated aperture magnitudes, taking advantage of SExtractor's dual-image mode to use the multiband image for source detection and the single-band images for measurement. We use an aperture radius of 1.3$''$ in both wavebands, and select objects according to the following criteria:
\begin{enumerate}
 \item stellarity $<$ 0.95 \& FWHM $>$ 0.2$''$, to avoid contamination from stars;
 \item magnitude uncertainty $<$ 0.2 mag \& flag $>$ 4, to avoid objects that may not have been properly deblended or do not have reliable photometry; 
 \item F625W magnitude $<$ 25 mag, as a luminosity cut-off.
\end{enumerate}
We cross-correlate our SExtractor catalogue with the existing CLASH catalogue, which provides photometric redshifts based on the full set of 16 ACS/HST filters, and further reject all objects whose redshift range do not satisfy $z_{min} < 0.545 < z_{max}$ for the 95\% confidence intervals $z_{min}$ and $z_{max}$.

We model the surface brightnesses of the galaxies in the F625W image using single-component S\'ersic profiles, with 
\begin{equation}
 I(R) = I_e \exp{{\Bigg[-k_n\Big(\frac{R}{R_e}\Big)^{\frac{1}{n}}- 1\Bigg]}}
\end{equation}
for S\'ersic index $n$, ffective radius $R_e$, $k_n = 2n - 0.324$ and radius $R^2 = q^2(x-x_0)^2 + (y-y_0)^2/q^2$ for axis ratio $q$ and galaxy centroid $(x_0,y_0)$ (with both $R_e$ and $R$ being circularised, projected quantities). To test for and eliminate systematics in our size measurements, we proceed via two different methods. First, we use the curve-of-growth fitting (COG) code presented in \citet{houghton}, which has been rigorously tested and shown to reproduce simulated images to high accuracy (see the appendix of that paper). This masks bright regions close to the galaxy to create a `clean' cutout, whose circularised integrated light profile is then fitted using a chi-squared minimisation. Second, we use a surface-fitting code based on that developed for \citet{Oldham2016b}, which explores the six-dimensional parameter space represented by $(x_0, y_0, R_e, n, q, \theta)$ -- where $(x_0,y_0)$ is the centroid of the S\'ersic profile, $\theta$ is the position angle and $q$ the ellipticity -- using the Markov Chain Monte Carlo (MCMC) sampler \texttt{emcee} \citep{ForemanMackey2013}. In this case, for objects with crowded fields we model all bright objects simultaneously. For both the COG and the surface fitting routines, we convolve the model with the radial profile of an unsaturated star in the image to account for the point-spread function (PSF), and we find very good agreement between the two sets of results. For Coma, we use the sizes and surface brightnesses presented in \citet{jorgensen95}, which were measured using de Vaucouleurs COGs.

Of the 31 MACSJ0717 objects in the spectroscopic sample, only 19 of these overlap with the CLASH field; of these, a further two are in fact stars (as can be clearly seen in both the imaging and the spectra), and so are excluded from the analysis. We therefore end up with a final sample of 17 galaxies. The completeness of the spectroscopic sample relative to the galaxy population is illustrated in Figure~\ref{fig:CMD}. Three of the systems in the final sample (object IDs 11, 13 and 22) have clear extended stellar haloes or bulge+disk morphologies which mean that we are unable to construct satisfactory models using a single S\'ersic component; for these, we add a second component and find that this enables us to model their light profiles down to the noise. It is important to do this sparingly in order that all our measured sizes are directly comparable.

The galaxy cutouts and model residuals are presented in Figure~\ref{fig:sersices} in the Appendix, and the effective radii and surface brightnesses are included in Table~\ref{tab:data}, with the latter corrected for cosmological dimming as $10\log(1+z)$ and extinction. Figure~\ref{fig:hst} shows the CLASH footprint, with our final galaxy sample marked. We also calculate synthetic rest-frame absolute magnitudes in the Johnson $U$ and $B$ bands, using the stellar population models of \citet{bruzual} (under the same assumptions as those stated in Section 4); these are included in Table A1.

\subsection{Spectroscopy}
\label{sub:spec}
We reduce the spectra of all 31 galaxies using the GMOS package in IRAF \citep{Tody1993}, calibrating the wavelength using the skylines in the exposures, according to the UVES sky emission atlas \citep{Hanuschik2003}. To attain a higher signal-to-noise, we stack the 12 exposures slitlet by slitlet. 

We model each spectrum as the sum of a galaxy and a continuum component. For the former, we use stellar templates for G, K, A and F stars from the Indo-US Stellar Library of Coud\'{e} Feed Stellar Spectra \citep{valdes}, which we redshift and convolve with a dispersion $\sigma_{model}^2 = \sigma_{true}^2 + \sigma_{inst}^2 - \sigma_{tmp}^2$ where $\sigma_{true}$ is the physical velocity dispersion of the system, $\sigma_{inst}$ is the instrument resolution and $\sigma_{tmp}$ is the intrinsic resolution of the templates (which is 1.2 $\mathrm{\AA}$ for the Indo-US templates). We measure the resolution in each spectrum by fitting Gaussians to the skylines, and find this to be constant across the slitlets, with $\lambda / d\lambda = 3030$. The continuum is an order-6 polynomial which accounts for the difference in shape between the templates and the true spectrum, and regions where atmospheric absorption dominates the spectrum are masked. We therefore have two free non-linear parameters -- the redshift and velocity dispersion for the galaxy -- and 15 linear parameters -- the weights of each of the nine stellar templates, and the coefficients of the order-6 polynomial. We construct a likelihood of the data given the model
\begin{equation}
 \ln L = -\frac{1}{2}\sum_k \Big(\frac{F_{k,obs} - F_{k,mod}}{\delta F_k}\Big)^2
\end{equation}
where the sum is over $k$ pixels along the wavelength axis, and $F_{mod}$, $F_{obs}$ and $\delta F$ are the model flux, observed flux and observed variance respectively. We then explore the posterior probability distribution of the model given the data using the Markov Chain Monte Carlo (MCMC) package \texttt{emcee} \citep{ForemanMackey2013}. 

Our kinematic models are shown in the far-right panels of Figure A1 and the resulting velocity dispersions are included in Table 1. We test the robustness of our kinematic inference by repeating the exercise using the lower-resolution galaxy templates of \citet{bruzual} and find that the typical uncertainty in the velocity dispersion is of order $5 \%$; as this is significantly larger than our statistical uncertainties, we impose this as the uncertainty on all our velocity dispersion measurements (though we note that the statistical uncertainties are given in Table 1). We also check the robustness of our method by modelling the spectra independently using the penalised pixel fitting (pPFX) software of \citet{cappellari04} (v4.15, also using the INDO-US library), and find the uncertainty to be less than $5 \%$. To our inferred velocity dispersions we apply aperture corrections following the prescription in \cite{jorgensen95}, correcting the dispersions from the $1''$ apertures over which we extracted the spectra to a standard aperture size of 3.4$''$ at the distance of Coma.


\begin{table*}
 \centering
\begin{tabular}{C{0.5cm}C{1.2cm}C{1.25cm}C{1.5cm}cccC{1.8cm}C{1.7cm}C{2.1cm}}\\\hline
 ID & RA/deg & DEC/deg & $R_e$/kpc & $\sigma$/kms$^{-1}$ & $\langle\mu_e\rangle$/mag & F625W/mag & F475W - F625W /mag & $\log (M_{\star}/M_{\odot})$ & $\log (M_{dyn}/M_{\odot})$ \\\hline
2 & 109.4079 & 37.7363 & $ 3.04 \pm  0.18 $ & $ 220.9 \pm 4.0 $ & $ 19.92 \pm 0.11 $ & $ 21.46 \pm 0.03 $ & $ 1.88 \pm 0.03 $ & $ 11.40 \pm 0.06 $ & $ 11.24 \pm 0.03 $ \\
4 & 109.3952 & 37.7316 & $ 3.42 \pm  0.20 $ & $ 198.8 \pm 5.4 $ & $ 19.93 \pm 0.10 $ & $ 21.24 \pm 0.04 $ & $ 1.86 \pm 0.03 $ & $ 11.42 \pm 0.09 $ & $ 11.20 \pm 0.03 $ \\
5 & 109.3943 & 37.7358 & $ 1.69 \pm  0.06 $ & $ 220.2 \pm 4.8 $ & $ 18.44 \pm 0.05 $ & $ 21.27 \pm 0.01 $ & $ 1.80 \pm 0.02 $ & $ 11.28 \pm 0.09 $ & $ 10.98 \pm 0.01 $ \\
7 & 109.3981 & 37.7515 & $ 4.06 \pm  0.12 $ & $ 298.7 \pm 3.0 $ & $ 19.57 \pm 0.05 $ & $ 20.49 \pm 0.05 $ & $ 1.89 \pm 0.02 $ & $ 11.73 \pm 0.11 $ & $ 11.63 \pm 0.02 $ \\
8 & 109.3986 & 37.7548 & $ 9.51 \pm  0.30 $ & $ 157.8 \pm 2.4 $ & $ 20.68 \pm 0.54 $ & $ 18.48 \pm 0.01 $ & $ 1.90 \pm 0.01 $ & $ 12.51 \pm 0.11 $ & $ 11.44 \pm 0.02 $ \\
9 & 109.3857 & 37.7454 & $ 14.55 \pm  0.33 $ & $ 274.7 \pm 4.7 $ & $ 21.37 \pm 0.02 $ & $ 19.54 \pm 0.01 $ & $ 1.70 \pm 0.01 $ & $ 11.61 \pm 0.15 $ & $ 12.11 \pm 0.01 $ \\
10 & 109.3892 & 37.7514 & $ 2.68 \pm  0.11 $ & $ 159.1 \pm 3.6 $ & $ 19.25 \pm 0.07 $ & $ 21.07 \pm 0.02 $ & $ 1.75 \pm 0.03 $ & $ 11.22 \pm 0.10 $ & $ 10.90 \pm 0.02 $ \\
11 & 109.3831 & 37.7535 & $ 2.52 \pm  0.97 $ & $ 184.8 \pm 3.9 $ & $ 20.51 \pm 0.09 $ & $ 20.59 \pm 0.03 $ & $ 1.77 \pm 0.02 $ & $ 11.42 \pm 0.10 $ & $ 11.00 \pm 0.03 $ \\
12 & 109.3707 & 37.7523 & $ 5.07 \pm  0.24 $ & $ 216.5 \pm 4.1 $ & $ 20.44 \pm 0.08 $ & $ 20.88 \pm 0.02 $ & $ 1.79 \pm 0.02 $ & $ 11.39 \pm 0.09 $ & $ 11.44 \pm 0.02 $ \\
13 & 109.3690 & 37.7577 & $ 2.41 \pm  0.88 $ & $ 109.4 \pm 3.1 $ & $ 20.66 \pm 0.08 $ & $ 20.84 \pm 0.05 $ & $ 1.69 \pm 0.03 $ & $ 11.08 \pm 0.15 $ & $ 10.52 \pm 0.03 $ \\
14 & 109.3601 & 37.7584 & $ 10.26 \pm  0.94 $ & $ 223.2 \pm 4.1 $ & $ 21.28 \pm 0.16 $ & $ 20.20 \pm 0.05 $ & $ 1.85 \pm 0.02 $ & $ 11.80 \pm 0.11 $ & $ 11.77 \pm 0.04 $ \\
16 & 109.3529 & 37.7672 & $ 11.31 \pm  0.42 $ & $ 302.8 \pm 3.8 $ & $ 20.82 \pm 0.05 $ & $ 19.52 \pm 0.04 $ & $ 1.89 \pm 0.02 $ & $ 11.99 \pm 0.12 $ & $ 12.08 \pm 0.02 $ \\
22 & 109.4019 & 37.7341 & $ 3.40 \pm  0.38 $ & $ 205.1 \pm 5.9 $ & $ 20.79 \pm 0.02 $ & $ 20.22 \pm 0.01 $ & $ 1.97 \pm 0.02 $ & $ 12.01 \pm 0.08 $ & $ 11.22 \pm 0.02 $ \\
23 & 109.3754 & 37.7511 & $ 3.33 \pm  0.13 $ & $ 227.1 \pm 6.7 $ & $ 19.88 \pm 0.05 $ & $ 21.25 \pm 0.02 $ & $ 1.92 \pm 0.03 $ & $ 11.57 \pm 0.07 $ & $ 11.30 \pm 0.01 $ \\
24 & 109.3629 & 37.7653 & $ 3.79 \pm  0.17 $ & $ 187.6 \pm 4.7 $ & $ 20.14 \pm 0.08 $ & $ 21.21 \pm 0.02 $ & $ 1.80 \pm 0.03 $ & $ 11.31 \pm 0.10 $ & $ 11.19 \pm 0.02 $ \\
28 & 109.4188 & 37.7439 & $ 0.21 \pm  0.00 $ & $ 166.1 \pm 5.2 $ & $ 17.52 \pm 0.18 $ & $ 24.77 \pm 0.14 $ & $ 1.34 \pm 0.03 $ & $ 8.51 \pm 0.06 $ & $ 9.84 \pm 0.01 $ \\\hline
\end{tabular}
\caption{Photometry, kinematics and stellar and dynamical masses for the galaxies in the final MACSJ0717 sample. The photometry has been corrected for extinction and effective surface brightnesses $\langle \mu_e \rangle$ are corrected for cosmological dimming as $10\log(1+z)$; $F625W$ magnitudes are calculated from surface fitting the HST/ACS image as described in Section 2.1, while $F475W-F625W$ colours are calculated from the SExtractor catalogues, also described in Section 2.1. Dynamical and stellar masses $M_{dyn}$ and $M_{\star}$ are calculated as described in Sections 3.3 and 3.4. We note that a number of objects have $M_{\star} > M_{dyn}$, which may be a result of the assumptions made to measure stellar masses, or the fact that $M_{dyn}$ is not the total dynamical mass but the mass measured in some ill-defined aperture. It also reflects the findings of a number of other studies of non-local ETGs \protect\citep[e.g.][]{Stockton2014}. We discuss this further in Section 5.3. We also compute synthetic rest-frame absolute magnitudes in the Johnson $U$ and $B$ bands; these are included in Table A1.}
\label{tab:data}
\end{table*}

\section{Scaling relations}
\label{sec:results}

\subsection{The Colour--Magnitude Relation}
\label{sub:CMR}

\begin{figure*}
\centering
\includegraphics[trim=20 10 10 10,clip,width=\textwidth]{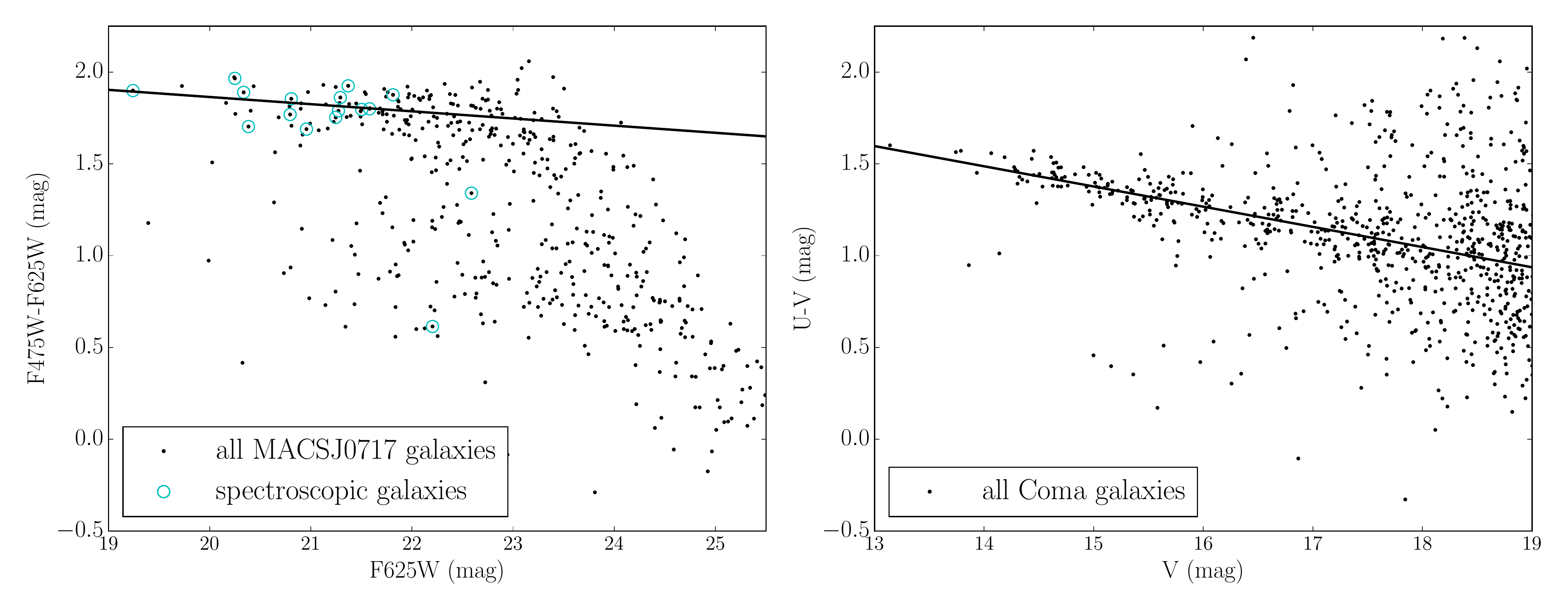}
\caption{The colour-magnitude relations for MACSJ0717 (left) and Coma (right), with all apparent magnitudes measured in the observed frame. The 17 galaxies included in the MACSJ0717 spectroscopic sample are circled in turquoise, and indicate that our spectroscopic subsample is 50\% complete down to an absolute magnitude of  -21.5 mag in the F625W band. The fits to the red sequences are also plotted.}
\label{fig:CMD}
\end{figure*}

We use a mixture model \citep{hogg10} to fit for the slope $\alpha_{CMD}$, intercept $\beta_{CMD}$ and intrinsic scatter $\sigma_{CMR}$ of the red sequence according to the equation

\begin{equation}
M_1 - M_2 = \alpha_{CMD} M_2 + \beta_{CMD} ,
\end{equation}
where $M_1$ and $M_2$ represent the `blue' and `red' magntiudes respectively, with $M_1, M_2 = U, V$ for Coma and $M_1, M_2 = F475W, F625W$ for MACSJ0717. Our model assigns every point a probability of belonging to either the linear distribution of the red sequence or a distribution of outliers that is Gaussian in colour, and seeks the best fit via an MCMC exploration. 

We test this routine by applying it to the data for Coma, and obtain a value for the scatter $\sigma = 0.065 \pm 0.009$, which agrees with the value $\sigma = 0.069 \pm 0.01$ quoted in \cite{terlevich01}. We note that, in that study, different morphological groupings of galaxies have their scatters measured separately and give significantly different results: however, in our sample we do not make any cuts based on morphology. Also, we find that our model is very sensitive  to the choice of upper magnitude limit, with stricter magnitude limits leading to smaller inferred scatters. This may be a sign of the inadequacy of a single Gaussian for describing the outlier distribution. To compare the scatter of the two clusters, then, it is important to cut both samples at the same physical magnitude, and it is the smaller sample of higher-redshift galaxies, with generally fewer galaxies at the faint end, that dictates where this should be. This sets an apparent magntiude cut-off for the higher-redshift cluster of $F625W = 24$ mag, which we convert to an $r$-band cut-off for the Coma galaxies using the stellar population models of Section 4.

We find the red sequence of MACSJ0717 to have a shallower slope than Coma, consistent with the fact that Coma is being observed in bluer filters. The scatter is consistent with that of Coma, though slightly smaller, indicating that they stellar populations are already old and red in the MACSJ0717 galaxies, as explored in Section 4.1 and discussed in Section 5.1. This may also be affected by the smaller interval sampled by the $U$ and $V$ filters in Coma's rest frame than by the MACSJ0717 filters. We also note that the MACSJ0717 galaxies appear to extend to brighter magnitudes than the Coma sample (the brightest Coma galaxy has an absolute magnitude $M_U = -20.2$, whereas the MACSJ0717 galaxies extend to $M_U = -19.42 \pm 0.18$, as shown in Table A1); though this is partly an effect of the different filters with which each cluster has been observed, the filters and redshifts are such that this would in fact be \textit{amplified} by a filter correction. This is a point that we return to in the anaylsis of other scaling relations in Section 3.3. The fitted red sequence of MACSJ0717 is shown in Figure~\ref{fig:CMD} and a summary of the results is given in Table~2.

\begin{figure*}
\centering
\subfigure{\includegraphics[trim=10 0 10 0,clip,width=0.5\textwidth]{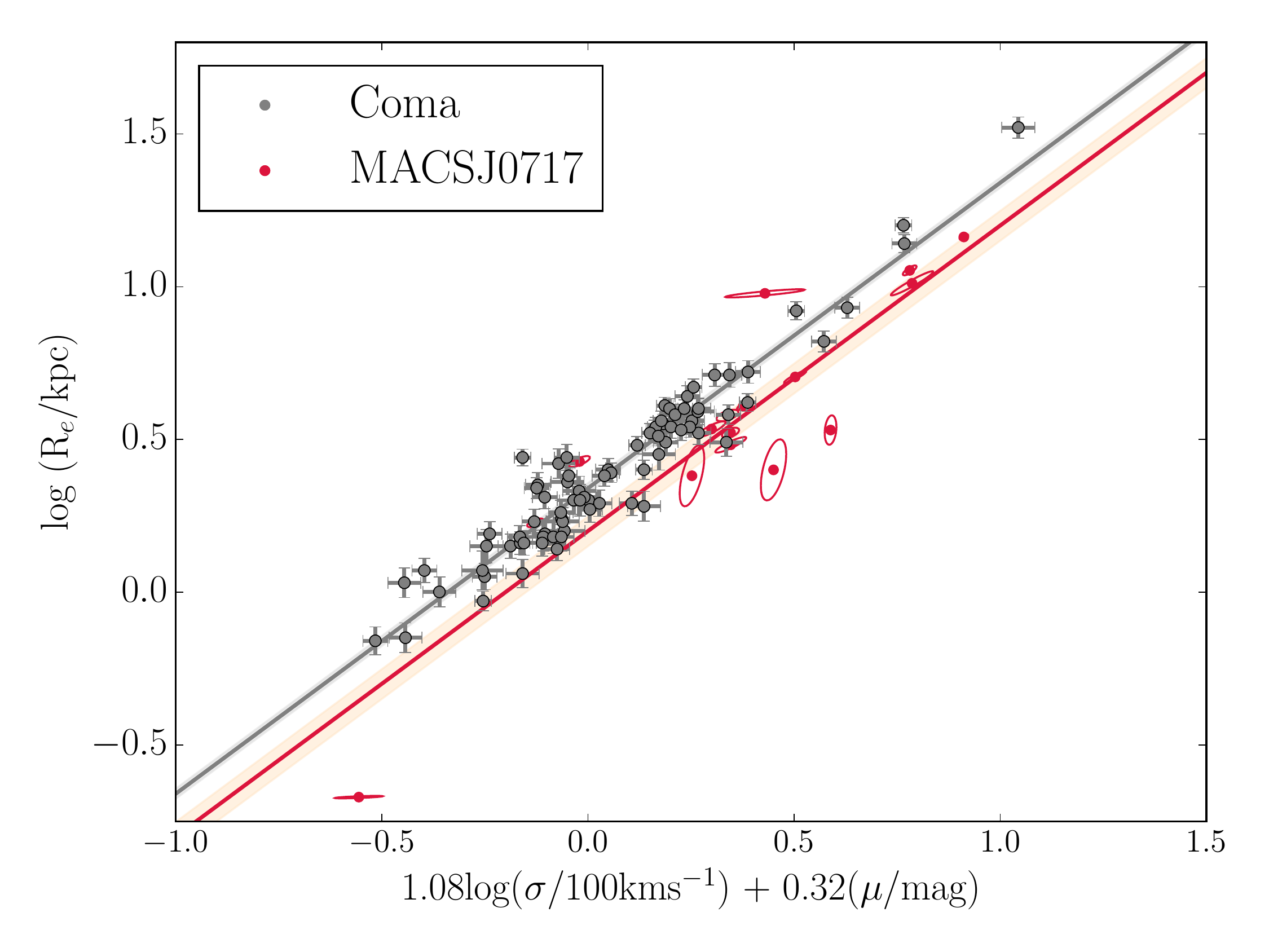}}\hfill
\subfigure{\includegraphics[trim=10 0 10 0,clip,width=0.5\textwidth]{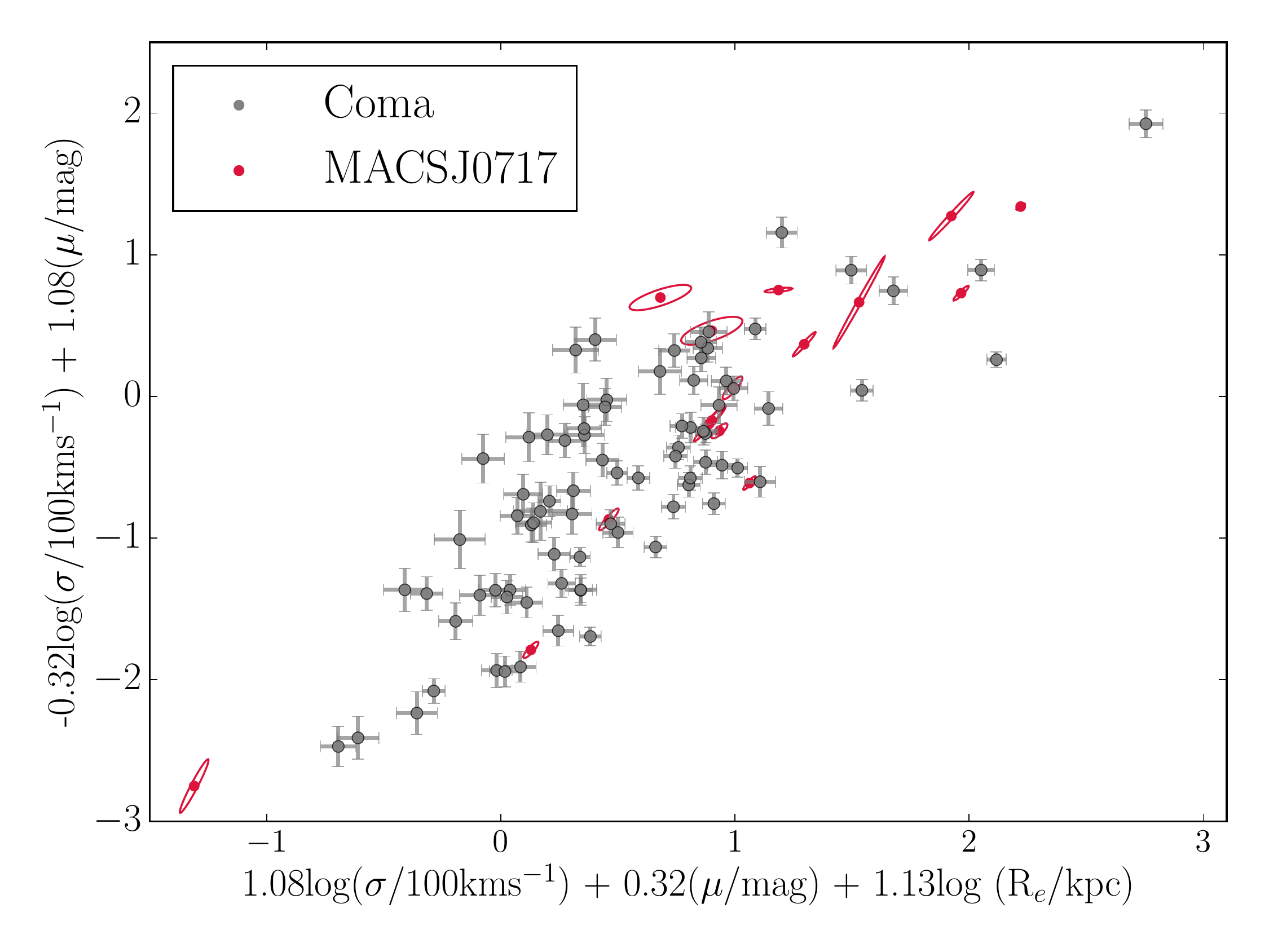}}\hfill
\caption{The FPs of the Coma and MACSJ0717 galaxies, with $\sigma$ in units of 100kms$^{-1}$, $\langle\mu_e\rangle$ in magnitudes and $R_e$ in kpc. The FP of MACSJ0717 is slightly offset from that of Coma, which we show to be the result of two effects: the evolution of the stellar populations and the evolution of the structure of the galaxies. Note that the surface brightnesses here are measured in the observed-frame F625W and $r'$ filters for MACSJ0717 and Coma respectively, and have been corrected for cosmological dimming as $10\log(1+z)$. Two complementary projections are shown in order to highlight the different regions occupied by the two galaxy populations. Left: An edge-on projection. Right: A face-on projection.}
\label{fig:FP}
\end{figure*}

\subsection{The Fundamental Plane}
\label{sub:FP}

Initially, we fit the FPs for the two clusters independently; as we find their slopes to be consistent with no evolution, we then model them simultaneously, requiring both to be parallel, in order to infer the offset between the two. We assume each dataset to lie on a plane

\begin{equation}
\label{eq:plane}
\begin{split}
 \log R_{e,z} = \alpha_{FP,z} \log \sigma_z + \beta_{FP,z} \langle \mu_{e,z} \rangle + \gamma_{FP,z} \\
 \log R_{e,0} = \alpha_{FP,0} \log \sigma_0 + \beta_{FP,0} \langle \mu_{e,0} \rangle + \gamma_{FP,0} \\
\end{split}
\end{equation}
where the subscript $x_0$ refers to quantities relating to the low-redshift cluster (Coma), and the subscript $x_z$ corresponds to the high-redshift cluster. To reduce the scale of the degeneracy between the plane parameters, we redefine $\sigma_{FP} = \sigma/100$kms$^{-1}$ and $\langle\mu_{e,FP}\rangle = \langle\mu_e\rangle - 20$. We further assume the independent variables to be drawn from a multivariate Gaussian distribution with mean $\vec{\nu}_{FP} = (\nu_{\log\sigma}, \nu_{FP,\mu_e})$ and variance 
\begin{equation}
 \vec{\tau}_{FP}^2 = \left( \begin{array}{cc}
\tau_{\log\sigma}^2 & \rho\tau_{\log\sigma}\tau_{\mu_e} \\
\rho\tau_{\log\sigma}\tau_{\mu_e} & \tau_{\mu_e}^2 \end{array} \right).
\end{equation}
This is appropriate as it deals with the fact that the galaxies in each sample are almost certainly drawn from different intrinsic distributions, with the higher-redshift MACSJ0717 galaxies likely to be drawn from the most massive and luminous end of their population. More details on the model are provided in \citet{Kelly2007}, which presents the formalism, but essentially, we construct the likelihood for the data given a particular set of plane and Gaussian parameters, and obtain the posterior distribution using an MCMC exploration of the parameter space. The median values of the marginalised distributions are presented in Table 3, together with 16th and 84th percentile uncertainties, and the fitted FP is shown in both edge-on and face-on projections in Figure~\ref{fig:FP}. We demonstrate that the MACSJ0717 galaxies are indeed drawn from a distribution with a higher mean velocity dispersion, with $\nu_{\log{\sigma}} = 0.31  \pm 0.03$ for MACSJ0717 and $\nu_{\log{\sigma}} = 0.21  \pm 0.02$ for Coma -- consistent with the idea that the former extend to brighter magnitudes and higher masses. When we model the two planes simultaneously, we find the clusters to be offset, with $\gamma_z - \gamma_0 = -0.14 \pm 0.06$.


\subsection{The $M_{dyn}/L-M_{dyn}$ relation}
\label{sub:ML}

To construct the MLM relation, we calculate the dynamical mass-to-light ratio within $R_e$, $M_{dyn}/L$, in the $F625W$ band (MACSJ0717) and $r$ band (Coma) using the virial estimator
\begin{equation}
 \frac{M_{dyn}}{L}(<R_e) = \frac{\beta\sigma^2R_e}{G L}
\end{equation}
\citep[e.g.][]{cappellari06}. This assumes virial equilibrium, consistent with the assumptions of the FP.  The value of the coefficient $\beta$ depends on the mass distribution of the galaxy, and can be either calculated from theoretical models (in which case it is dependent on the S\'ersic index $n$) or calibrated from galaxy samples for which multiple $M_{dyn}/L$ estimates are available. As our size measurements for the Coma galaxies come from de Vaucouleurs models, we do not have S\'ersic indices from which to calculate $\beta(n)$ (though we do for the MACSJ0717 galaxies); we therefore adopt the best-fitting value of $\beta = 5$ presented in \cite{cappellari06}, which was calibrated by comparing virial and Schwarzschild $M_{dyn}/L$ estimates for a sample of 25 E and S0 galaxies. We then construct a `dynamical' mass
\begin{equation}
 M_{dyn} = \frac{\beta\sigma^2R_e}{G}
\end{equation}
though we note that this is not intended to represent the total dynamical mass within any physically meaningful aperture. For instance, according to Equation 6, the half-light dynamical mass should use $\beta = 2.5$, whereas \citet{Wolf2010} suggests that $\beta = 4$ is more appropriate. Clearly, there remains uncertainty here, and we therefore continue to use $\beta = 5$ in order to facilitate comparisons with other studies. Given that we are interested in the offset between the two galaxy samples rather than the absolute relations, the choice of any constant $\beta$ does not affect our conclusions (though this is not true if $\beta$ varies across the plane).

\begin{figure}
\centering
\includegraphics[trim=20 0 10 0,clip,width=0.5\textwidth]{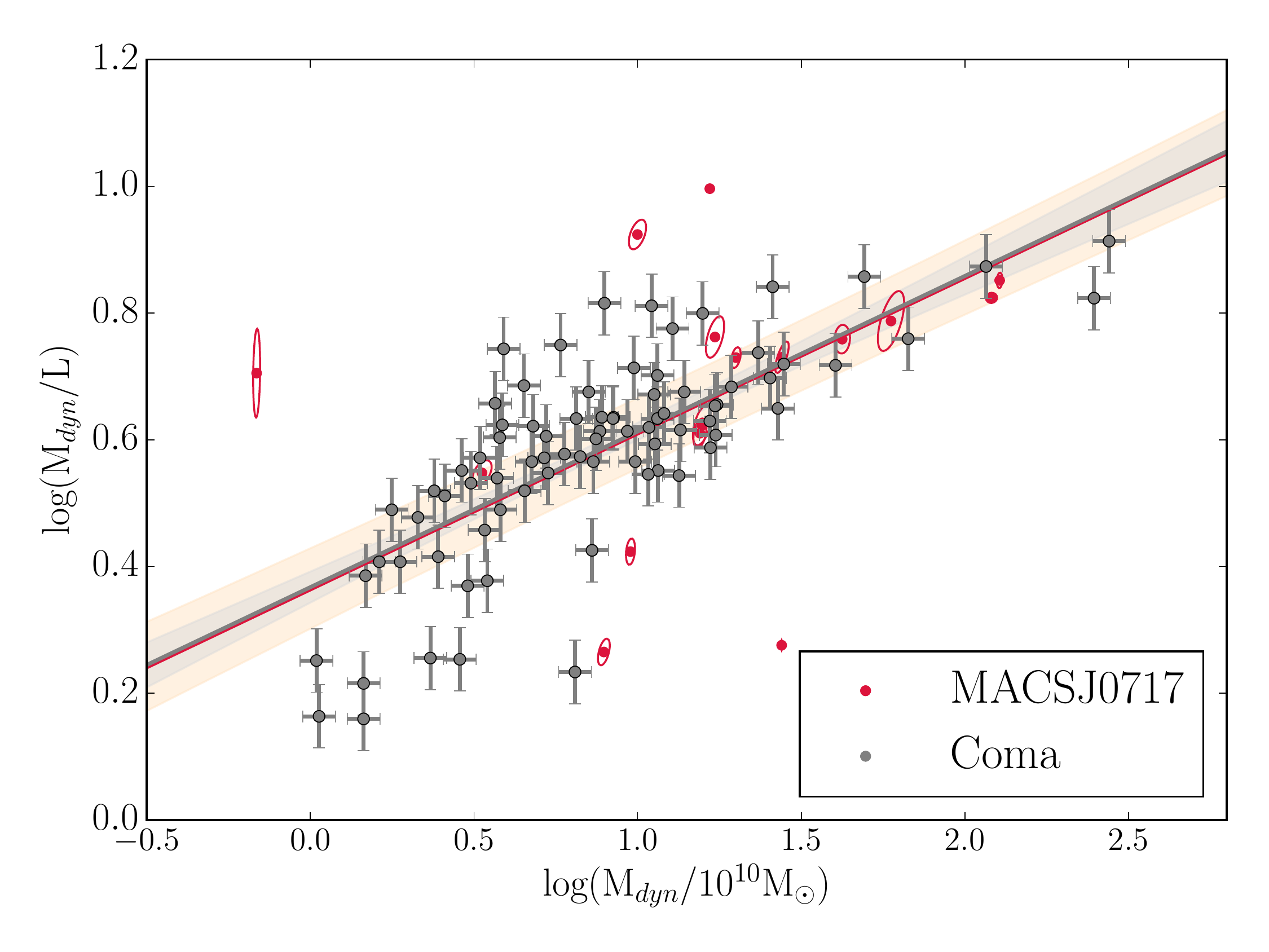}
\caption{The MLM relation for the galaxies in Coma and MACSJ0717, with $M_{dyn}$ in units of $10^{10} M_{\odot}$ and $M_{dyn}/L$ in units of $M_{\odot}/L_{\odot}$. When the MACSJ0717 galaxies are constrained to be parallel to those of Coma, the relations lie virtually on top of each other (though note that the luminosity is measured in a different filter and at a different redshift for each cluster).}
\label{fig:ML}
\end{figure}

We calculate the bandpass luminosities relative to that of the Sun (based on a redshifted CALSPEC solar spectrum) from the surface brightnesses using the equation
\begin{equation}
 \mathcal{M}_{gal} - \mathcal{M}_{\odot} = -2.5\log\Big(\frac{L_{gal}}{L_{\odot}}\Big)
\end{equation}
where $\mathcal{M}$ represents an absolute magnitude and $\mathcal{M}_{\odot}(z=0,r) = 4.58$ and $\mathcal{M}_{\odot}(z=0.545,F625W) = 5.17$, and construct the MLM relations for the Coma and MACSJ0717 galaxies, as can be seen in Figure~\ref{fig:ML}. Again modelling the masses and mass-to-light ratios as being drawn from a multivariate Gaussian distribution, we infer both these underlying distributions and the slope, intercept and scatter of the linear relation

\begin{equation}
 \log \frac{M_{dyn}}{L} = \alpha_{ML} \log M_{dyn} + \beta_{ML}.
\end{equation}
where mass $M_{dyn}$ and luminosity $L$ are measured in units of $10^{10}M_{\odot}$ and $L_{\odot}$; our inference is presented in Table 2.
This time, modelling the clusters independently leads us to find marginally different slopes (in addition to an offset), with $\beta_{ML} = 0.25 \pm 0.02$ for Coma and $\beta_{ML} = 0.12 \pm 0.11$ for MACSJ0717, though the uncertainties of the MACSJ0717 relation are large, making them consistent at the $1\sigma$ level. The point here is that our high-redshift sample lacks the dynamic range that would be needed to robustly infer both the slope and intercept of the MLM relation; these two parameters suffer degeneracies, making them hard to constrain. When we then model the two populations together, we find that the Coma data dominate the fit to the slope -- which is not surprising, given that we have $\sim$ 6 times more galaxies in the latter -- such that $\beta_{ML} = 0.25 \pm 0.02$; the MLM relation for MACSJ0717 then lies virtually on top of that of Coma, with $\beta_{z} - \beta_{0} = -0.01 \pm 0.06$. We also confirm that the MACSJ0717 galaxies are drawn from a more massive distribution, with $\nu_{ML} = 1.24 \pm 0.15$ compared to $\nu_{ML} = 0.88 \pm 0.05$ for Coma. This is a selection effect that we would expect, given that the former is at a higher redshift and that magnitude limits mean that we are only able to observe the most massive end of the mass distribution. Nevertheless, Figure 5 suggests there are genuinely more high-mass galaxies in MACSJ0717 than in Coma -- this is an interesting result that may be connected with the higher cluster mass of MACSJ0717 (see Section 5.4). Figure~\ref{fig:ML} also shows more generally that the two populations have similar trends between their dynamical masses and mass-to-light ratios, with the main difference being that the MACSJ0717 galaxies have higher masses. The results of this modelling are summarised in Table 2.

\subsection{The $M_{\star}-\sigma$ and $M_{\star}-R_e$ relations}
\label{sec:size}

As discussed in the Introduction, any structural evolution of the galaxy population with redshift would also have an impact on the offset of the FP. We therefore attempt to measure the evolution in size and velocity dispersion between the two ETG samples using two independent methods. First, we fit the $M_{\star}-R_e$ and $M_{\star}-\sigma$ relations to infer the difference in size and velocity dispersion of galaxies of any given mass; that is the subject of this Section. Later, we model the FP and MLM, $M_{\star}-R_e$ and $M_{\star}-\sigma$ relations simultaneously in order to infer the evolution in size, velocity dispersion and luminosity all at once. That is the topic of Section 4.5.

We use stellar masses here as opposed to the dynamical masses calculated in the previous Section to avoid the obvious degeneracies between $M_{dyn}$, $\sigma$ and $R_e$.
We calculate stellar masses for Coma cluster and MACS0717 galaxies by comparing the $u-g$ and F475W-F625W colours, respectively, to the same colours calculated from the SSP models of \citet[BC03]{bruzual}. In order to break the degeneracy between age and metallicity, we make the assumption that galaxies on the red sequence are coeval to first-order and that the slope of the red sequence is driven by a systematic change in metallcity with luminosity \citep[as found by][]{kodama}. By assuming an \emph{average} age of the red sequence, we interpolate between SSP models of fixed age and varying metallicity to convert the colour-magnitude relation of the red sequence into a metallicity-luminosity relation; this provides a metallicity for each galaxy, based on its luminosity and not its colour. With this metallicity, we use the observed colour to infer the age and stellar mass-to-light ratio $M_{\star}/L$ of each galaxy. In practice, for each galaxy we interpolate between the two SSPs that bracket the metallicity derived from the CMR to calculate the age and $M_{\star}/L$. We do not apply any luminosity weighting corrections as these are deemed second order. Nor do we limit the derived ages to be younger than the age of the Universe (although this is not a significant issue).
 
For young star forming galaxies, $M_{\star}/L$ is far smaller, and the luminosity far greater, than that of an old passive galaxy of the same mass, leading us to dramatically overestimate of the metallicity from the luminosity-metallicity relation we derived above for red sequence galaxies. However, at young ages ($<1$Gyr), colour is primarily determined by age, not metallicity. Thus curves of colour versus $M_{\star}/L$ for different metallicities only diverge at old ages; at younger ages, the curves converge and the dominant factor in determining $M_{\star}/L$ is age; metallicity has virtually no effect. Thus for luminous blue star forming galaxies, although our luminosity-metallicity relation overestimates the metallicity, the $M_{\star}/L$ and stellar masses remain accurate. Hence we find no need to iterate the estimation of the metallicity once the age and $M_{\star}/L$ (the real parameters of interest) have been calculated.

Adopting this method, we calculate the $M_{\star}/L$ for the Coma galaxies assuming red sequence ages of  8, 10, \& 12 Gyrs. We then adopt the average of the $M_{\star}/L$ values and use the scatter from the variation of input age as the formal error on the $M_{\star}/L$ values. We find Coma cluster galaxies to be around solar metallicity and (by construction) around 10 Gyrs old. When calculating the $M_{\star}/L$ of the galaxies in MACS0717, we adopt red sequence ages of (8, 10, 12) - 5.3 Gyrs. We find that the MACSJ0717 galaxies are slightly higher metallicity (around 1.5 $\times$ solar) than Coma galaxies (around solar). Note that although the derived stellar masses depend slightly on the initial age assumed for Coma (and for MACS0717, forced to be 5.3 Gyrs younger), the relative ratio between the Coma and MACS0717 masses is almost constant if we assume an age $>8$ Gyrs for Coma. For ages below 8 Gyrs, the stellar masses for MACSJ0717 galaxies drop rapidly compared to the stellar masses for Coma galaxies (to the extent that if we assume an age of 6 Gyrs for Coma, the average stellar masses in both clusters become the same). But such young ages (implying ages of $<$1 Gyr for the MACSJ0717 stellar populations) are unrealistic and ruled out by both the CMR and FP or MLM results. 

The stellar masses are included in Table 2; we note that the stellar masses reveal the general mass differences of the two samples even more clearly than the dynamical masses of the previous Section, with almost all the high-$z$ galaxies containing more stellar mass than nearly all the low-$z$ galaxies. As discussed previously, this indicates a genuine excess of high-mass galaxies in MACSJ0717 relative to Coma. 

\begin{figure*}
 \centering
\subfigure{\includegraphics[trim=10 0 10 10,clip,width=0.48\textwidth]{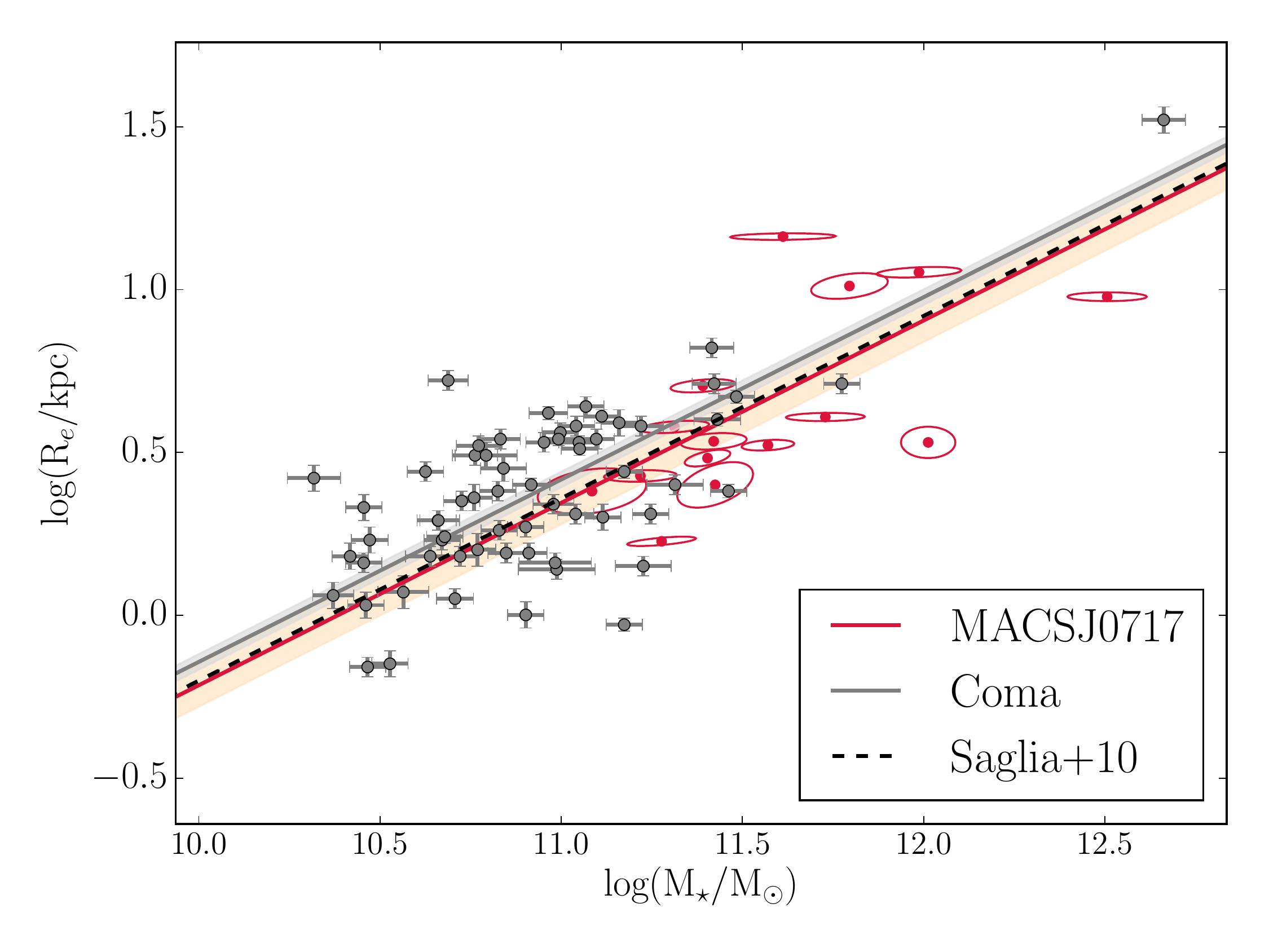}}\hfill
\subfigure{\includegraphics[trim=10 0 10 10,clip,width=0.48\textwidth]{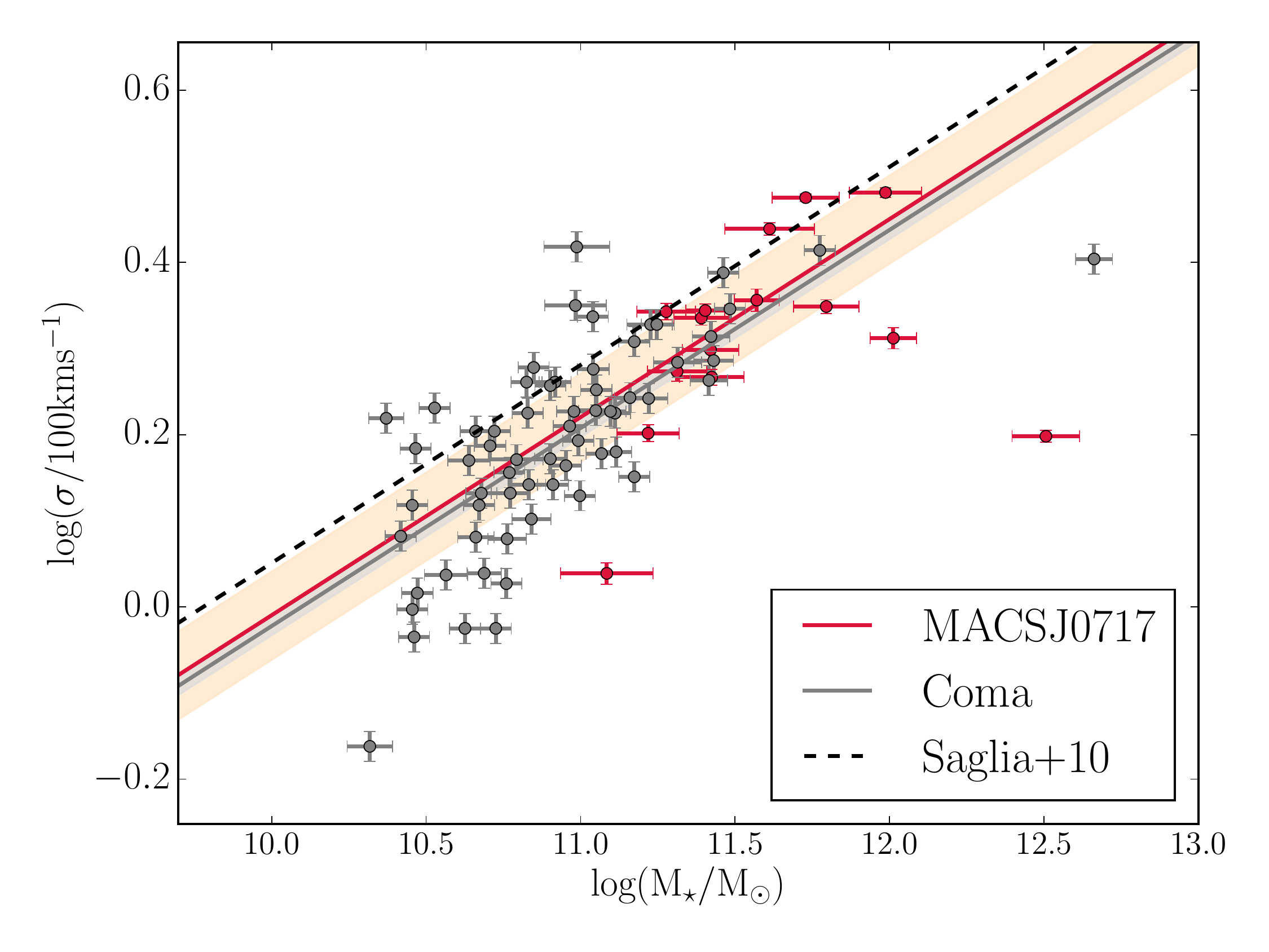}}\hfill
\caption{Mass-size and mass-velocity dispersion relations for the two clusters, with stellar masses in units $10^{10} M_{\odot}$, sizes in kpc and velocity dispersions in units of 100 kms$^{-1}$. In both cases, the scaling relations of the two clusters are consistent with being parallel but offset, with the high-redshift galaxies being both marginally smaller and having marginally higher velocity dispersions. The dashed lines show the relations given in \protect\citet{saglia10}, evaluated at $z=0.545$. Both are consistent with our MACSJ0717 sample within $2\sigma$ (though we do not show the uncertainties on the \protect\citealp{saglia10} relations here).}
\label{fig:mstars}
\end{figure*}

We assume the size and velocity dispersion to follow power laws in the total mass and fit
\begin{equation}
 \log R_e = \alpha_{MR} \log M_{\star} + \beta_{MR}
\end{equation}
and
\begin{equation}
 \log \sigma = \alpha_{MS} \log M_{\star} + \beta_{MS}
\end{equation}
using the same formalism as in earlier Sections in which the slope, intercept and intrinsic scatter and inferred together with the properties of the underlying Gaussian distribution of $\log M_{\star}$. The limited dynamic range in mass of each population makes it difficult to make meaningful inference on both the slope and intercept of these relations (especially for the size-mass relation) and break the strong degeneracy that exists between them: we therefore fix the slope of the size-mass relation to $\alpha_{MR} = 0.56$, as found in \citet{Shen2003}, and that of the sigma-mass relation to $\alpha_{MS} = 0.23$ as in \citet{saglia10}, both of which were measured using significantly bigger galaxy samples. Our inference is shown in Figure~\ref{fig:mstars} and demonstrates that the two populations look extremely similar in both respects. This implies that a very small amount of evolution has taken place between the $z \sim 0.5$ and $z \sim 0$ galaxies.

We also compare our $M_{\star}-R_e$ and $M_{\star}-\sigma$ relations with those of \citet{saglia10} -- in which these relations were constructed for galaxies from 26 clusters out to redshifts $z\sim0.9$ -- and find that we are consistent with both at the $2 \sigma$ level. With regard to the slightly poorer agreement between the $M_{\star}-\sigma$ relations relative to the $M_{\star}-R_e$ relations, we note that MACSJ0717 has a high velocity dispersion $\sigma_{cluster} = 1660_{-130}^{+120}$ kms$^{-1}$ \citep{ebeling07} relative to the mean velocity dispersion $\bar{\sigma}_{cluster} = 525 \pm 210$ kms$^{-1}$ of the EDisCS clusters used in that study. This may be evidence for the more rapid evolution of galaxies in denser environments, such that the galaxy velocity dispersions in higher-mass clusters at $z=0.5$ more closely resemble those of $z=0$ galaxies than do those in lower-mass clusters at $z=0.5$. We discuss this further in Section 5.2, though we cannot make any strong claims on the basis of the data used in this paper.

Following \citet{VanderWel2008} and \citet{saglia10}, we relate the offsets between the clusters to measure the evolution as a function of redshift:
\begin{equation}
 \beta_{MR,z} - \beta_{MR,0} = \xi \log (1+z)
\end{equation}
and
\begin{equation}
 \beta_{MS,z} - \beta_{MS,0} = \eta \log (1+z)
\end{equation}
and therefore find $\xi = -0.37 \pm 0.39$, $\eta = 0.06 \pm 0.28$ as summarised in Table 2. We note that the evolution we find here is weaker than the $\xi = -0.98 \pm 0.11$ found by \citet{VanderWel2008} for field ellipticals, but consistent with the $\xi = -0.53 \pm 0.04$ inferred by \citet{Delaye2014} for cluster ellipticals, and moreover that the evolution in both size and velocity dispersion is consistent with zero \citep[see also][for indications of mild structural evolution of cluster ETGs]{saglia10,Jorgensen2014,Saracco2014}. The implication is that only a small amount of evolution has taken place in these galaxies between $z = 0.545$ and the present day. We use these in Section 4.4 to account for the effects of size evolution in the FP and MLM relations.

\section{Stellar Population Models}
\label{sec:SPS}

The key idea of using the changes in our scaling relations to understand the underlying stellar populations is that the scaling relations themselves are simply a convenient way of characterising the observable properties of galaxies - in our case, this means luminosities, colours, sizes, velocity dispersions - and that these properties are governed in turn by the stars they contain. We can then use stellar population synthesis (SPS) models, under particular, astrophysically-motivated assumptions regarding the age, metallicity, IMF and SFH of the population, to connect the changes we observe to the evolution of the stellar populations. The assumptions we make as to the SFHs in the CMR and the FP and $M_{dyn}/L$ analyses are different and are simplifications of the real, much more complex and extended processes that we know ETGs are subject to: however, both are motivated by the data, and can be interpreted together to provide a fuller picture of these galaxies' evolution. The other main assumption we make here is that the ETGs in MACSJ0717 are directly comparable to those in Coma, such that the former represent the Coma population at an earlier stage in their evolution.

Historically, the differences due to the redshifts of the two galaxy populations and the filters in which they have been observed have been accounted for by \textit{correcting the data}; specifically, by applying K-corrections to the data and then comparing the FP zeropoints as if both galaxy populations had been observed at the same redshift and with the same filter. However, this requires some galaxy `template' to be chosen and assumptions to be made regarding the spectral energy distribution (SED) and age of the galaxies, and therefore introduces signficant uncertainty and possible bias. We therefore refrain from doing this, and, rather than correcting the data, \textit{we entirely forward-model the observations}. In a development of the methods introduced in \citet{houghton}, we use the BC03 SPS models, assuming a Salpeter IMF \citep[based on evidence that massive ETGs may have IMFs that more heavier than the Milky-Way-like Chabrier IMF, e.g.][]{Auger2010} and solar metallicity, and entirely forward-model the data by evaluating the SPS models in the same redshifted filters as those with which the latter were observed (though we do subtract the cosmological dimming term $10\log(1+z)$ from the surface brightnesses, as stated in Section 2.2). This removes the need to apply any ad-hoc colour or bandpass `corrections' to the data \citep[see e.g.][]{Hogg2002} based on assumptions about the SED. Any further assumptions that we make regarding the stellar populations in the case of specific scaling relations are explained in the relevant section.


\subsection{Luminosity evolution from the CMR}

To constrain the stellar ages from the evolution in the CMR scatter, we assume an SFH in which each galaxy comprises a simple stellar population (SSP), but allow for a spread in SSP ages between galaxies. This allows us to write the colour scatter as a Taylor expansion 
\begin{equation}
 \frac{\mathrm{d}\textrm{col}}{\mathrm{d}t} \approx \sigma_{CMR} / \sigma_{age} \approx 3.5 \sigma_{CMR} / \Delta t
\end{equation}
where $\frac{\mathrm{d}\textrm{col}}{\mathrm{d}t}$ is the rate of change of colour at the mean stellar age of the galaxies, $\sigma_{CMR}$ is the intrinsic colour scatter measured from the CMR, $\sigma_{age}$ is the intrinsic scatter in stellar age between galaxies (assuming a Gaussian distribution) which translates to the equivalent width of $\Delta t \approx 3.5\sigma_{age}$ of a population of galaxies that form their SSPs uniformly between $t_{start}$ and $t_{stop}$. Relating $t_{start}$ and $t_{stop}$ by the ratio $b$ of the SSP formation period to the total time available, 
\begin{equation}
\Delta t = t_{stop} - t_{start} = b t_{stop},
\end{equation}
we can rewrite the mean formation time $t_f = \frac{1}{2}(t_{start}+t_{stop})$ in terms of $\Delta t$ and so derive an equation for the evolution in colour of a galaxy's stars:
\begin{equation}
\frac{\mathrm{d}\textrm{col}}{\mathrm{d}t} = 3.5\sigma _{CMR} \frac{b ^{-1} - \frac{1}{2}}{t_f}
\label{eq:dcol}
\end{equation}
\citep[see also][]{bower92,bower98,houghton}. Note that this model has all the stars in a single galaxy forming simultaneously, but distributes the formation times for different galaxies uniformly with a dispersion $b$ and mean age $t_f$. Thus $b = 0$ corresponds to a cluster whose galaxies all formed at once, whereas $b = 1$ allows the cluster galaxies to have formed their stars from the beginning of the Universe until $t_{stop}$.

Given that we have inferred the intrinsic scatter $\sigma_{CMR}$ for the CMRs of the two clusters, and that we can use SPS models to calculate the rate of change of colour as a function of stellar age, we can thus infer the stellar age dispersion $b$ and mean stellar age $13.6 - t_f$ Gyr of the cluster galaxies. We do this by constructing a chi-squared likelihood from Equation 16, accounting for the uncertainties in $\sigma_{CMR}$ that are given in Table 2. As shown in Figure 6, we find a mean age $9.44_{-0.57}^{+0.46}$ Gyr and dispersion $b = 0.83_{-0.12}^{+0.11}$, indicating that the stellar populations are fairly old but formed with a significant dispersion $\sim 3$ Gyr. Note that this is a strong constraint compared to the lower limits that have been previously obtained from the CMR \citep{bower92,bower98,houghton}; this is mainly due to the fact that we have measured the intrinsic scatter for two clusters  rather than one, and can therefore break the degeneracy between $b$ and $t_f$, which would otherwise be unconstrained. The strength of our constraint relative to \citet{houghton}, who did use two clusters, comes from the wider redshift separation of our clusters. The constraint obtained here could therefore be further improved by the addition of higher-redshift clusters to this analysis.

\begin{figure}
\centering
\includegraphics[trim=20 20 20 20,clip,width=0.49\textwidth]{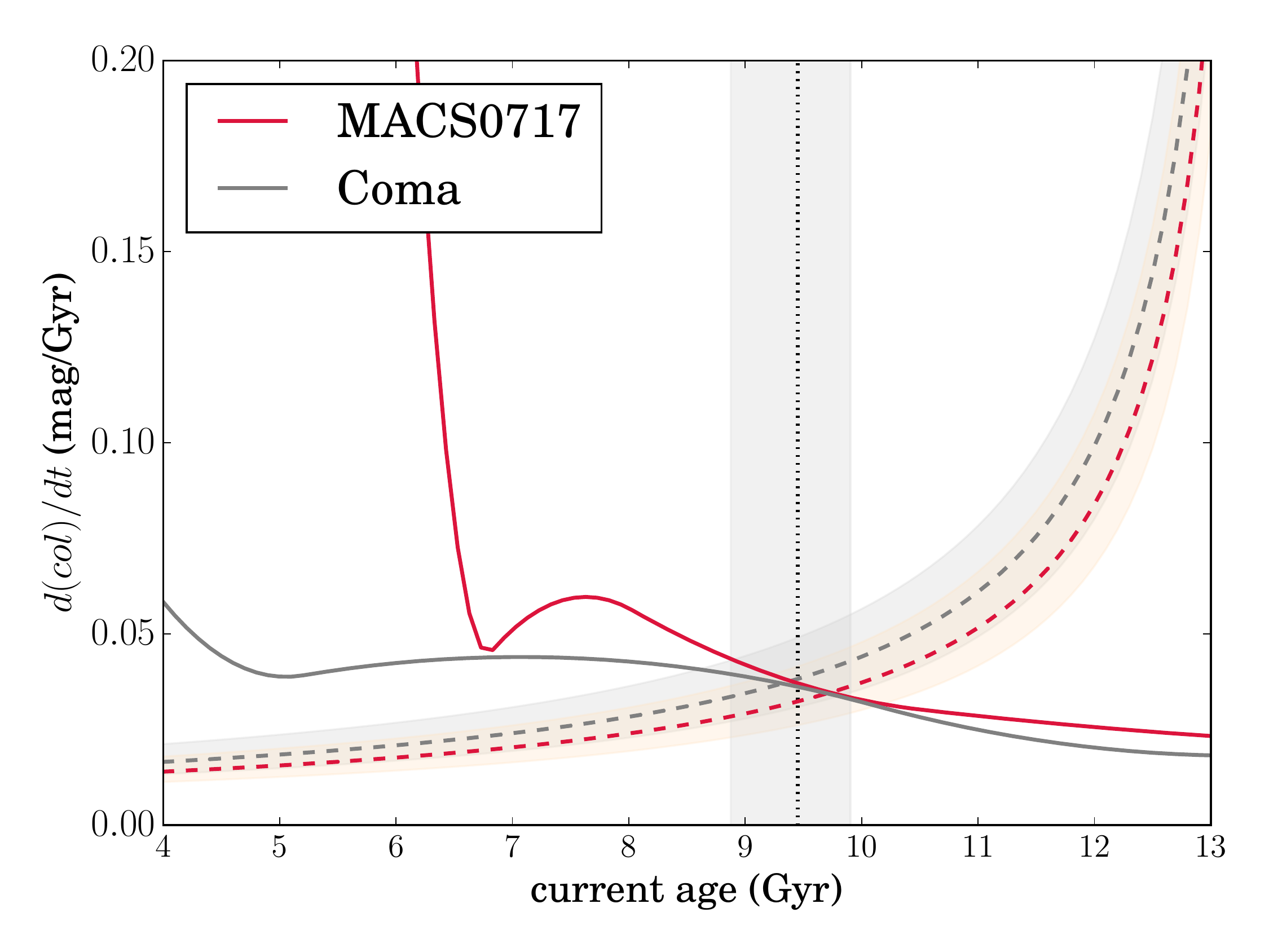}
\caption{Inferring the mean stellar age and age dispersion of the cluster galaxies from the evolution of the CMR's intrinsic scatter. Assuming each galaxy to consist of an SSP, with a spread formation times between galaxies, we compare the implied rate of change of colour with that predicted by SSP models. The solid curves show the rate of change of colour that for a solar metallicity SSP with a Salpeter IMF in the BC03 models as a function of the current age of the stars; the dashed curves show the right-hand side of Equation 16 for the value $b = 0.83_{-0.12}^{+0.11}$ that we infer, and the vertical black line shows the inferred mean age $9.44_{-0.57}^{+0.46}$ Gyr.}
\label{fig:age}
\end{figure}

\subsection{Luminosity evolution from the fundamental plane}
\label{sub:lum}

We have observed that there is a change in the zero-point of the fundamental plane with time. This could be due to an evolution in any or all of the plane variables - indeed, at fixed mass, all are understood to change with redshift, though the evolution in $R_e$ and $\sigma$, as discussed in Section 1, is understood to be a small effect. Similarly to \cite{saglia10} and \cite{VanderWel2008}, we derive a general expression relating the FP offset to the evolution of the FP variables, and apply this, first under the assumption of no size evolution, and then using a parameterisation for size evolution which we constrain directly from our data. In contrast to the CMR analysis, we now assume that all the stars in all the galaxies formed in a single event, with one SSP characterising the whole cluster. We can then compare the luminosity evolution we observe with the predictions of SPS models to obtain a further estimate of the stellar age. 

By requiring that the FP of MACSJ0717 be parallel to that of Coma, we have been able to measure its displacement, $\Delta\gamma_{FP} = \gamma_{FP,z} - \gamma_{FP,0} = -0.14 \pm 0.06$, in the direction of $\log R_e$, where subscripts $x_0$ and $x_z$ denote quantities measured with respect to the low- and high-redshift clusters respectively, as before. Given the FP equation and the construction of the surface brightness as $\langle\mu_e\rangle = -2.5\log(\frac{L}{2\pi R_e^2})$, we have
\begin{equation}
 \gamma_z - \gamma_0 = \Delta\log R_e - \alpha\Delta\log\sigma - \beta\Delta\langle\mu_e\rangle
\end{equation}
where we have dropped the subscript $x_{FP}$ for clarity, and $\Delta X = X_z - X_0$, i.e. $\Delta X > 0$ means that $X$ has decreased between redshift $z$ and today. This translates into a magnitude evolution $\Delta m$
\begin{equation}
\label{eq:offset1}
\Delta m = \frac{(1-5\beta)\Delta\log R_e - \alpha\Delta\log\sigma - \Delta\gamma}{\beta}.
\end{equation}
In Section~4.5, we use the full machinery of Equation~\ref{eq:offset1} to both investigate and treat the effects of size evolution. Here, though, we proceed under the first approximation that $R_e$ and $\sigma$ are constant with redshift, in which case Equation~\ref{eq:offset1} simplifies dramatically. 

The measured change between $\gamma_{z}$ and $\gamma_{0}$ translates to an evolution in magnitude of $\Delta m = 0.44 \pm 0.10$ mag (in the different filters), which we include in Table~\ref{tab:ML}. This implies that, for a particular position on the FP, the stars in a MACSJ0717 galaxy are actually \textit{dimmer} than those in Coma. While, at first glance, this appears contrary to our expectation of ageing stellar populations, we emphasise that the high- and low-redshift filters are \textit{not} matched, meaning that each is being sampled in a different region of the spectrum. The purpose of our SPS model comparison is to account for these filter effects in addition to the effects of the intrinsic luminosity evolution of the population. 

We use the BC03 stellar population models to interpret this, again assuming an SSP with solar metallicity and a Salpeter IMF. As explained at the beginning of Section 4, we take account of both the different filters used and the age difference by modelling the colour $\Delta m = F625W(z=0.545,T-5.33)-r(z=0,T)$, where $z$ is the redshift and $T$ is the age of the Universe in Gyr. The evolution of $\Delta m$ in this setup is shown in Figure 7, with the magnitude offset $\Delta m = 0.44 \pm 0.19$ determined from the FP overplotted in red, along with its 1$\sigma$ upper and lower bounds. The intersection of this measured offset with the model lines gives an estimate for the age as $9.12_{-0.80}^{+1.22}$ Gyr, and is in good agreement with the constraints from the CMR that were obtained in the previous Section, plotted in grey on the figure. This is old, and implies that the stellar populations are already highly evolved by the time we observed them in MACSJ0717.

\begin{figure}
\centering
\includegraphics[trim=20 10 10 20,clip,width=0.5\textwidth]{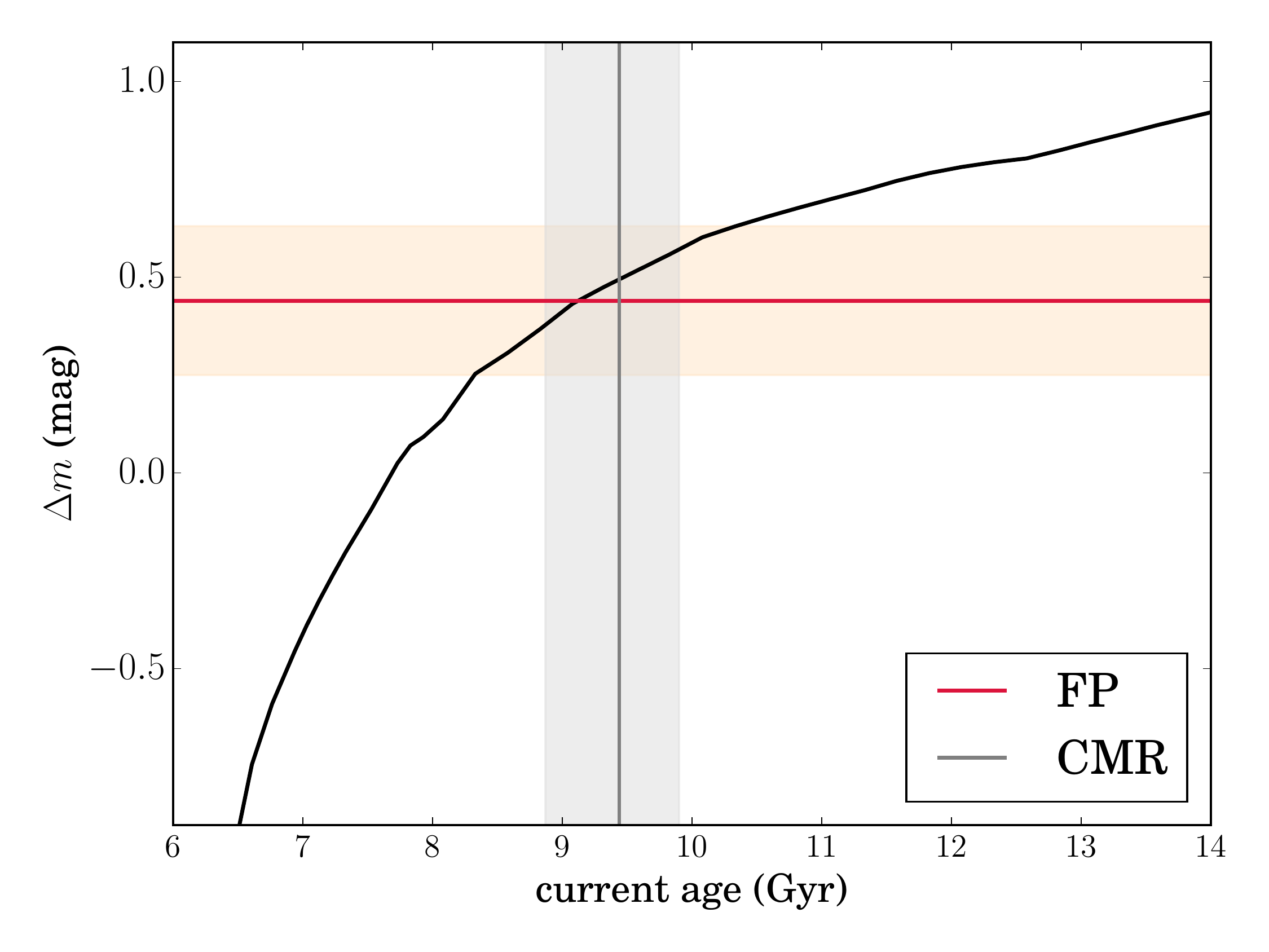} 
\caption{SSP prediction for the evolution of the observed magnitude difference between the Coma and MACSJ0717 galaxies as a function of cluster age. Overplotted in red is the offset inferred from the evolution of the FP, as described in Section 4.2; the grey vertical line shows the inference on age from the CMR, which agrees remarkably well. In both cases, the shaded regions show the $1 \sigma$ unccertainties.}
\label{fig:mu}
\end{figure}

\subsection{Luminosity evolution from the $M_{dyn}/L-M_{dyn}$ relation}
\label{sec:mlm}

As for the FP, we assume that both populations can be described  by the same power-law index (the individual cluster slopes are consistent at the $1 \sigma$ level) and, under the same assumption of one universal SSP, we use our simultaneous fit to the two clusters and compare the offset between them with that predicted by SSP models in order to make another estimate of the stellar age. Here, given that the mass scales as $M \sim \sigma^2 R_e$, we have
\begin{equation}
 \Delta \log (M/L) = 2 \Delta \log \sigma + \Delta \log R_e - \Delta \log L
\end{equation}
which can be related to the offset using Equation 9 to give
\begin{equation}
 \Delta m = \Delta \mathcal{M_{\odot}} + 2.5\big[ \Delta \beta_{ML} + (\alpha_{ML} - 1)(2 \Delta\log \sigma + \Delta\log R_e)\big]
\end{equation}
where $\Delta\mathcal{M_{\odot}}$ is the difference in absolute solar magnitude in the (blueshifted) filters. 

As with the FP, we initially assume zero structural evolution, and calculate the magnitude offset and corresponding stellar age. From our best fit to the two cluster populations, we then find $\Delta m = 0.57 \pm 0.18$, consistent with our earlier FP result and included in Table 4. This implies a stellar age of $9.89_{-0.98}^{+1.73}$ Gyr.

\subsection{Combining size and luminosity evolution}

To apply our inference on size evolution from the mass-size and mass-velocity dispersion relations of Section 3.4 to our scaling relations using Equations 16 and 18, we define

\begin{equation}
 \Delta \log R_e = \xi \log(1+z)
\label{eq:xi}
\end{equation}
and
\begin{equation}
 \Delta \log \sigma = \eta \log(1+z),
\label{eq:eta}
\end{equation}
using the values of $\xi$ and $\eta$ as defined in Equations 11 and 12 and tabulated in Tables 2 and 3. 

Using our calculated values for $\xi$ and $\eta$ alongside equations 16 and 18, we can now calculate the magnitude evolution $\Delta m$ that must have taken place between the two clusters according to the FP and the MLM relation. We can write the magnitude offset as

\begin{equation}
 \Delta m = \frac{\log(1+z) \Big[(1-5\beta)\xi - \alpha \eta\Big] - \Delta \gamma}{\beta}
\end{equation}
for the FP, and
\begin{equation}
 \Delta m = \Delta\mathcal{M_{\odot}} + 2.5 \Big[ \Delta\beta + (\alpha-1)(2\eta+\xi)\log(1+z)\Big]     
\end{equation}
for the MLM relation. 

In Table 4, we summarise $\eta$, $\xi$, $\Delta m$ and the corresponding stellar age, according to our stellar population models, both with and without allowing for structural evolution. The final two columns of that Table give the implied magnitude evolution in the rest-frame Johnson $B$ and $V$ bands, assuming the stellar age that has been inferred in each case. We note that we do not attempt to account for progenitor bias that may arise from the fact that relatively young galaxies included in the local sample would be missing at higher redshifts, where they would not yet appear passive; we do not have the data to constrain it here. However, \citet{Valentinuzzi2010} and \citet{saglia10} investigated this for their larger samples of field galaxies at comparable redshifts, and found the effect to be small. We therefore assume the same to be true for our cluster galaxy samples. The MLM relation generally implies ages that are between 0.5 and 1 Gyr older than the FP, though the results are formally consistent and imply a stellar age of $\sim 10$ Gyr.

\begin{table*}
 \centering
\begin{tabular}{C{1.4cm}C{1.6cm}C{1.4cm}C{1.4cm}C{1.4cm}C{1.4cm}C{1.5cm}C{1.5cm}C{1.4cm}}\hline 

 parameter & CMR & FP & FP (parallel) & MLM & MLM (parallel) & $M_{\star}-\sigma$ (parallel) & $M_{\star}-R_e$ (parallel) & ALL \\\hline
$\alpha_{CMR}$ & $-0.03 \pm 0.01$ & --& --& --& --& --& --& --\\
$\beta_{CMR}$ & $1.17 \pm 0.01$ & --& --& --& --& --& --& --\\
$\sigma_{CMR}$ & $0.055 \pm 0.009$ & --& --& --& --& --& --& --\\\hline

$\alpha_{FP}$ & --& $1.15 \pm 0.43$ & $1.08 \pm 0.07$ & --& --& --& --&$1.10 \pm 0.18$ \\
$\beta_{FP}$ & --& $0.34 \pm 0.05$ & $0.32 \pm 0.01$ &--& --& --& --&$0.34 \pm 0.03$ \\
$\gamma_{FP}$ & --& $0.18 \pm 0.15$ & $0.20 \pm 0.05$ & --& --& --& --& -- \\
$\nu_{FP,\log\sigma}$ & --& $0.31 \pm 0.03$ & $0.31 \pm 0.03$ & --& --& --& --&$0.31 \pm 0.03$ \\
$\nu_{FP,\mu_e}$ & --& $0.10 \pm 0.28$ & $0.08 \pm 0.30$ & --& --& ---& --&$0.10 \pm 0.29$ \\
$\tau_{FP,\log\sigma}$ & --& $0.12 \pm 0.03$ & $0.12 \pm 0.03$ &--& --& --& --&$0.12 \pm 0.03$ \\
$\tau_{FP,\mu_e}$ & --& $1.13 \pm 0.25$ & $1.15 \pm 0.25$ & --& --& --& --&$1.15 \pm 0.26$ \\
$\rho$ & --& $0.11 \pm 0.25$ & $0.12 \pm 0.27$ & --& --& --& ---&$0.12 \pm 0.26$ \\
$\sigma_{FP}$ & --& $0.17 \pm 0.05$ & $0.16 \pm 0.04$ & --& ---& --& --&$0.12 \pm 0.06$ \\\hline

$\alpha_{ML}$ & --& --& --& $0.12 \pm 0.11$ & $0.25 \pm 0.02$& ---& --& $0.25 \pm 0.04$ \\
$\beta_{ML}$ & --& --& --& $0.52 \pm 0.15$ & $0.36 \pm 0.06$& --& --&-- \\
$\nu_{ML}$ & --& --& --& $1.24 \pm 0.15$ & $1.24 \pm 0.15$& ---& --& $ 1.25 \pm 0.16$ \\
$\tau_{ML}$ & --& --& --& $0.60 \pm 0.14$ & $0.59 \pm 0.14$& ---& --& $0.60 \pm 0.14$ \\
$\sigma_{ML}$ & --& --& --& $0.22 \pm 0.05$ & $0.22 \pm 0.05$& --& --& $0.22 \pm 0.05$ \\\hline

$\beta_{MR}$ & --& --& --& --& --& -- & -$0.21 \pm 0.07$ & --\\
$\nu_{MR}$ & --& --& --& --& --& -- & $1.39 \pm 0.22$ & $1.39 \pm 0.15$\\
$\tau_{MR}$ & --& --& --& --& --& -- & $0.90 \pm 0.20$ & $0.86 \pm 0.13$\\
$\sigma_{MR}$ & --& --& --& --& --& -- & $0.24 \pm 0.06$ & $0.24 \pm 0.06$\\\hline

$\beta_{MS}$ & --& --& --& --& --& -$0.01 \pm 0.05$ & -- & --\\
$\nu_{MS}$ & --& --& --& --& --& $1.39 \pm 0.23$ & -- & $1.39 \pm 0.15$\\
$\tau_{MS}$ & --& --& --& --& --& $0.92 \pm 0.20$ & -- & $0.86 \pm 0.13$ \\
$\sigma_{MS}$ & --& --& --& --& --& $0.20 \pm 0.05$ & -- & $0.20 \pm 0.04$\\\hline

$\xi$ & --& --& --& --& --& -- & -$0.37 \pm 0.39$& -$0.40 \pm 0.32$\\
$\eta$ & --& --& --& --& --& $0.06 \pm 0.28$& --& $0.09 \pm 0.27$\\
$\Delta m$ & --& --& --& --& --& --& --& $0.59 \pm 0.26$\\\hline

\end{tabular}
\caption{The inferred parameters for the CMR, FP, MLM, MR and MS relations of MACSJ0717, modelled as described in Section 3.}
\end{table*}

\begin{table*}
 \centering
\begin{tabular}{C{1.4cm}C{1.6cm}C{1.5cm}C{1.5cm}C{1.4cm}C{1.4cm}C{1.5cm}C{1.5cm}C{1.5cm}}\hline 

 parameter & CMR & FP & FP (parallel) & MLM & MLM (parallel) & $M_{\star}-\sigma$ (parallel) & $M_{\star}-R_e$ (parallel) & ALL \\\hline
$\alpha_{CMR}$ & -$0.11 \pm 0.01$  & --& --& --& -- --& -- \\
$\beta_{CMR}$ & -$0.82 \pm 0.01$  & --& --& --& --& --& -- \\
$\sigma_{CMR}$ & $0.065 \pm 0.005$  & --& --& --& --& --& -- \\\hline

$\alpha_{FP}$ & --& $1.09 \pm 0.07$ & $1.08 \pm 0.07$ & --& --& --& --&$1.10 \pm 0.18$ \\
$\beta_{FP}$ & --& $0.32 \pm 0.01$ & $0.32 \pm 0.01$ &--& --& --& --& $0.34 \pm 0.03$ \\
$\gamma_{FP}$ & --& $0.34 \pm 0.02$ & $0.34 \pm 0.02$ & --& --& --& --&$0.35 \pm 0.04$ \\
$\nu_{FP,\log\sigma}$ & --& $0.21 \pm 0.02$ & $0.21 \pm 0.02$ & --& --& --& --&$0.19 \pm 0.02$ \\
$\nu_{FP,\mu_e}$ & --& -$0.48 \pm 0.09$ & -$0.48 \pm 0.09$ & --& --& --& --&-$0.53 \pm 0.10$ \\
$\tau_{FP,\log\sigma}$ & --& $0.14 \pm 0.01$ & $0.14 \pm 0.01$ &--& --& --& --&$0.12 \pm 0.01$ \\
$\tau_{FP,\mu_e}$ & --& $0.78 \pm 0.07$ & $0.78 \pm 0.06$ & --& --& --& --&$0.82 \pm 0.08$ \\
$\rho$ & --& -$0.09 \pm 0.11 $ & -$0.09 \pm 0.11$ & --& --& --& --&-$0.24 \pm 0.13$ \\
$\sigma_{FP}$ & --& $0.07 \pm 0.01$ & $0.08 \pm 0.01$ & --& --& --& --&$0.05 \pm 0.03$ \\\hline

$\alpha_{ML}$ & --& --& --& $0.25 \pm 0.04$ & $0.25 \pm 0.02$& --& --& $0.25 \pm 0.04$ \\
$\beta_{ML}$ & --& --& --& $0.36 \pm 0.04$ & $0.37 \pm 0.02$& --& --&$0.37 \pm 0.03$ \\
$\nu_{ML}$ & --& --& --& $0.88 \pm 0.05$ & $0.88 \pm 0.06$& --& --& $ 0.82 \pm 0.05$ \\
$\tau_{ML}$ & --& --& --& $0.49 \pm 0.04$ & $0.48 \pm 0.04$& --& --& $0.39 \pm 0.04$ \\
$\sigma_{ML}$ & --& --& --& $0.07 \pm 0.01$ & $0.10 \pm 0.01$& --& --& $0.10 \pm 0.01$ \\\hline

$\alpha_{MR}$ & --& --& --& --& --& --& -- & -- \\
$\beta_{MR}$ & --& --& --& --& --& -- & -$0.14 \pm 0.03$ & -$0.14 \pm 0.02$\\
$\nu_{MR}$ & --& --& --& --& --& -- & $0.93 \pm 0.05$ & $0.93 \pm 0.04$\\
$\tau_{MR}$ & --& --& --& --& --& -- & $0.38 \pm 0.04$ & $0.38 \pm 0.03$\\
$\sigma_{MR}$ & --& --& --& --& --& -- & $0.19 \pm 0.02$ & $0.19 \pm 0.02$\\\hline

$\alpha_{MS}$ & --& --& --& --& --& -- & -- & --\\
$\beta_{MS}$ & --& --& --& --& --& -$0.02 \pm 0.01$ & -- & -$0.02 \pm 0.01$\\
$\nu_{MS}$ & --& --& --& --& --& $0.93 \pm 0.05$ & -- & $0.93 \pm 0.04$\\
$\tau_{MS}$ & --& --& --& --& --& $0.38 \pm 0.04$ & -- & $0.38 \pm 0.03$\\
$\sigma_{MS}$ & --& --& --& --& --& $0.08 \pm 0.01$ & -- & $0.08 \pm 0.01$\\\hline

$\xi$ & --& --& --& --& --& -- & -$0.37 \pm 0.39$& -$0.40 \pm 0.32$\\
$\eta$ & --& --& --& --& --& $0.06 \pm 0.28$& --& $0.09 \pm 0.27$\\
$\Delta m$ & --& --& --& --& --& --& --& $0.59 \pm 0.26$\\\hline

\end{tabular}
\caption{The inferred parameters for the CMR, FP, MLM, MR and MS relations of Coma, modelled as described in Section 3.}
\end{table*}

\begin{table*}
\caption{Inferences on the size, magnitude and velocity dispersion evolution from the FP, MLM and $M-R_e$ and $M-\sigma$ relations and the CMR, and the implied formation times of the stellar populations. All models recover a stellar age $\sim 10$ Gyr, though the MLM relation implies slightly larger ages than the FP and the joint analysis. The final two columns provide the implied magnitude evolution in the rest-frame Johnson $B$ and $V$ bands, according to the ages given in the sixth column.}
\label{tab:ML}
\centering
\begin{tabular}{cC{1.5cm}ccC{2.5cm}c|C{2.5cm}C{2.5cm}}\hline

scaling relation & size evolution corrected? & $\xi$ & $\eta$ & observed-frame magnitude evolution / mag & age / Gyr & rest-frame $B$-band magnitude evolution / mag & rest-frame $V$-band magnitude evolution / mag \\\hline

CMR & -- & -- & -- & -- & $9.44_{-0.57}^{+0.46}$ & -- & -- \\

FP & N & -- & -- & $0.44 \pm 0.19$ & $9.12_{-0.80}^{+1.22}$ & $-0.86_{-0.16}^{+0.15}$ & $-0.78_{-0.16}^{+0.14}$  \\

FP & Y  &  -$0.37 \pm 0.39$& $ 0.06 \pm 0.28$ & $0.54 \pm 0.29$ & $9.72_{-1.40}^{+3.17}$ & $-0.78_{-0.24}^{+0.21}$ & $-0.69_{-0.23}^{+0.18}$  \\
 
MLM & N & -- & -- & $0.57 \pm 0.18 $ & $9.89_{-0.98}^{+1.73}$ & $-0.75_{-0.15}^{+0.14}$ & $-0.67_{-0.15}^{+0.12}$ \\ 

MLM & Y  & -$0.37 \pm 0.39 $& $ 0.06 \pm 0.28$ & $0.66 \pm 0.19$ & $10.64_{-1.34}^{+2.50}$ & $-0.69_{-0.15}^{+0.13}$ & $-0.61_{-0.14}^{+0.11}$ \\
 
MLM \& FP & Y & -$0.40 \pm 0.32$ & $0.09 \pm 0.27$ & $0.59 \pm 0.26$ & $10.00_{-1.32}^{+3.14}$ & $-0.74_{-0.09}^{+0.19}$ & $-0.65_{-0.08}^{+0.15}$ \\\hline 

\end{tabular}
\end{table*}

\subsection{Combining scaling relations: inferring size and luminosity evolution}

In the previous sections, we analysed each scaling relation separately, using assumptions the size evolution measured from the $M_{\star}-R_e$ (MR) and $M_{\star}-\sigma$ (MS) relations to infer the evolution in luminosity and so the stellar age. Here, we model the FP and the MLM, MR and MS relations simultaneously, using both clusters so as to infer not only the scaling relation parameters and the underlying (Gaussian) distributions as before, but also to infer the magnitude, size and dynamical evolution between $z = 0.545$ and the present day. This has a number of significant advantages, including that (a) it ensures that the inferred scaling relations are all consistent, (b) it allows us to infer the physical parameters $\xi$, $\eta$ and $\Delta m$ (which were previously calculated after the modelling) in addition to those describing the scaling relations themselves and (c) it fully explores degeneracies between the physical parameters, as is not possible to do when they are calculated post-modelling.

To do this, we use the same formalism as before. We assume the MLM relations for the two clusters can be described according to the following equations

\begin{equation}
 \log \frac{M_{dyn}}{L} = \alpha_{ML} \log M_{dyn} + \beta_{ML} (z)
\end{equation}
where the intercept for the low-redshift cluster is
\begin{equation}
 \beta_{ML} (0) = \beta_{ML}
\end{equation}
and for MACSJ0717 we now explicitly account for size, magnitude and velocity dispersion evolution with
\begin{equation}
\beta_{ML} (z) = \beta_{ML} + 0.4(\Delta m - \Delta\mathcal{M_{\odot}}) + (1-\alpha_{ML})(2\eta + \xi)\log (1+z) .
\end{equation}
As before, we take the slope $\alpha_{ML}$ to be the same for both clusters. Also as before, each cluster has a distribution in $\log M_{dyn}$ given by a normal distribution with mean $\nu_{ML}$ and variance $\tau_{ML}^2$. 

We also assume the FPs for the two clusters can be described according to 

\begin{equation}
 \log R_e = \alpha_{FP} \log \sigma + \beta_{FP} \langle\mu_e\rangle + \gamma_{FP} (z)
\end{equation}
where the intercept of the low-redshift cluster is
\begin{equation}
 \gamma_{FP} (0) = \gamma_{FP}
\end{equation}
and for MACSJ0717
\begin{equation}
 \gamma_{FP} (z) = \gamma_{FP} + \Big[ \xi(1 - 5\beta_{FP}) - \alpha\eta\Big]\log(1+z) - \beta_{FP} \Delta m.
\end{equation}
Again, the slopes $\alpha_{FP}$ and $\beta_{FP}$ are the same for both clusters, and the independent variables $\log \sigma$ and $\langle\mu_e\rangle$ are drawn from multivariate normal distributions as described in Section 3.2.

Finally, we include the MR and MS relations, retaining our definition of $\eta$ and $\xi$ as being measured at constant stellar mass, to give
\begin{equation}
 \beta_{MR}(z) = \beta_{MR,0} + \xi \log(1+z) 
\end{equation}
and 
\begin{equation}
 \beta_{MS}(z) = \beta_{MS,0} + \eta \log(1+z) 
\end{equation}
for the MR and MS relations respectively, with $\log M_{\star}$ being drawn from a normal distribution with mean $\nu_{M_{\star}}$ and variance $\tau_{M_{\star}}^2$. This model now has the advantage of allowing us to infer the amount of size evolution and magnitude evolution that best describe our whole dataset, and guarantees that all four scaling relations are treated in a consistent way. It also sidesteps some of the potential dangers of our earlier method for constraining the structural evolution, as it does not assume that the stellar mass remains constant with redshift.

Our results are summarised in Table 2: encouragingly, the parameters of the FP and MLM relations are consistent with those inferred in our previous, simpler models. However, we now have additional constraints on the evolution of $R_e$, $\sigma$ and the luminosity. Although $\xi$, $\eta$ and $\Delta m$ have degeneracies within each relation, the modelling of all four relations at once breaks this degeneracy and we are able to infer $\Delta m = 0.59 \pm 0.26$, $\xi = -0.40 \pm 0.32$ and $\eta = 0.09 \pm 0.27$. The fact that the uncertainties on these parameters are comparable to -- and, in a number of cases, smaller than -- the uncertainties on the same parameters when each scaling relation is modelled separately, indicates that the degeneracies are not signficant, and that the scheme we have set up is indeed internally consistent. These results correspond to a stellar age $10.00_{-1.32}^{+3.14}$ Gyr.

\section{Discussion}
\label{sec:discussion}

\subsection{Old, passively evolving stellar populations}
\label{sec:ages_discussion}

The FP and its evolution with redshift contain a wealth of information about ETG formation and evolution; however, in order to extract this information meaningfully, it is important to understand the contributions due to different processes -- that is, the luminosity evolution of the stellar populations and the structural evolution of the galaxies themselves -- and to find a way to disentangle them. In this work, we have combined the FP with a number of other scaling relations in order to break these degeneracies and make inference on both the luminous and structural evolution. We are now in a position to tie together what we have found.

Initially, we used the evolution of the intrinsic scatter of the CMR to infer the mean stellar age, allowing for some dispersion. There, we assumed an SFH in which each galaxy is composed of an SSP with some dispersion in age across the galaxy population, and used the small evolution in the intrinsic scatter to infer a mean age $9.44_{-0.57}^{+0.46}$ Gyr and a dispersion of $\sim 3$ Gyr. This dispersion is significant but still implies some coordination in the star formation times of the different galaxies; together with the small intrinsic scatter of the FP and the MLM relation, this justified our treatment of the latter assuming that all the galaxies' stars formed in a single burst. We then modelled the FP, MLM, MS and MR relations in two ways: first, treating each separately, constraining the slopes to be parallel for the two clusters and using the offsets between them to measure the evolution in size, velocity dispersion and luminosity and hence the stellar age; second, by requiring all four relations to have evolved in a consistent way with regard to the structures and luminosities of the galaxies. In both cases, we find very clearly that only a small amount of evolution has taken place, with the high-redshift galaxies only marginally smaller and with marginally higher velocity dispersions than the Coma galaxies, and the luminosity evolution consistent with the passive fading of old populations.

In Table 4, we present the magnitude evolution that was inferred in each case, both in the observed-frame $r'$ (Coma) and F625W (MACSJ0717) filters and the rest-frame $U$ and $V$-band filters (though we note that the latter are more uncertain due to assumptions made in calculating K-corrections). In the joint analysis, we find $\Delta m_{B} = -0.74_{-0.09}^{+0.19}$ mag, or equivalently, $\Delta \log M/L_{B} = -0.30_{-0.04}^{+0.08} = -0.55_{-0.07}^{+0.15} z = -1.59_{-0.21}^{+0.42} \log(1+z)$. This is consistent with the findings of \citet{saglia10} and \citet{holden10}, indicating that the stellar populations in these ETGs have been evolving passively. Our inferred mean age of $10.00_{-1.32}^{+3.14}$ Gyr, corresponding to a formation redshift $z_{form} = 1.87_{-0.58}^{+>10}$, is also in agreement with the measurements of \citet{Jorgensen2014}, which examined the FP of a $z=1.27$ cluster, and implies that these galaxies are dominated by old stars which formed $\sim 10$ Gyr ago with some dispersion. Thus, we are seeing galaxies which are already significantly evolved when we look at MACSJ0717, consistent with a picture in which massive ETGs form their stars early and then grow passively and dissipationlessly, e.g. by minor mergers and accretion.

\subsection{Accelerated growth?}

The extremely small amount of structural evolution that we find to have taken place indicates that the galaxies in MACSJ0717 must have undergone the majority of their structural changes at earlier times. This may be a result of the very dense environment in which they are residing: indeed, other studies of galaxies in rich clusters out to $z \sim 1$ have also found no evidence for significant size or velocity dispersion evolution \citep{Stott2011, JorgensenChiboucas2013, Jorgensen2014, Saracco2014}. Moreover, when we compare these results with those from similar studies focussing on galaxies in lower-density clusters, in which stronger structural evolution is found \citep[e.g.][]{saglia10} -- and further, with those from studies of field ellipticals, which show evidence for yet stronger evolutionary trends \citep[e.g.][]{VanderWel2008,vanderWel2014} -- a tentative picture emerges of an environment-dependent growth timescale, with galaxies in denser environments reaching their present-day sizes at earlier epochs than those in lower-density environments. Whilst we cannot comment quantititavely on this hypothesis based on the data in this paper, we note that this would also be in line with the majority of studies that have directly  compared the sizes of passive galaxies in high- and low- density environments \citep[e.g.][]{Lani2013,Delaye2014} and found the galaxies in higher-density environments to be up to 50\% bigger \citep[though see also][]{Newman2014}.

If it is indeed the case that the growth of these galaxies has been accelerated by the dense cluster environment, this would also be strong evidence in favour of merger-driven growth -- which is likely to be enhanced in clusters -- as opposed to growth by internal processes such as adiabatic expansion due to quasar outflows \citep[e.g.][]{Fan2010}. It would therefore be interesting to take a deeper census of the MACSJ0717 cluster in order to establish whether the implied rate of mergers and accretion is consistent with the rate of evolution that we have observed. Of course, it is possible that our spectroscopic galaxy sample in MACSJ0717 is biassed towards the largest-radius systems, in which case the cluster may also host a number of other massive but smaller galaxies which are still undergoing some structural evolution. While a deeper census would again be necessary before this could be ruled out, it is nevertheless clear that a significant population of large, massive, apparently fully evolved galaxies are already in place by $z \sim 0.5$. We also note that MACSJ0717 is an extremely massive cluster -- indeed, the CLASH survey exclusively targeted strong lensing clusters -- and that it may therefore be an extreme example of accelerated growth.

\subsection{Can we trust the stellar and dynamical masses?}

The stellar masses derived in Section 3.4 are on average higher than the dynamical masses, for the galaxies in both clusters (see Table 1) -- implying that all the mass in these systems should be luminous. For Coma, the median ratio of stellar to dynamical mass is 1.39, while for MACSJ0717 the median ratio is 1.48. At face value this is unphysical. However, recall that the dynamical mass given by Equation 7 is not the total dynamical mass, but is really twice the dynamical mass within one effective radius. Furthermore, with $\beta=5$, it calculates the dynamical mass for a specific mass profile. In fact, variations in $\beta$ may be as large as a factor of two for typical mass profiles of ETGs. Furthermore, when calculating the mass-to-light ratios from the BC03 stellar populations, we adopted a Salpeter IMF. Mass-to-light ratios in the $r$ band for old, solar-metallicity SSPs are typically in the ratio 2:3 for Chabrier : Salpeter IMFs. Thus had we adopted a \citet{Chabrier} or \citet{Kroupa} IMF, the stellar masses would be roughly equal to the dynamical masses. We further note that we are not the first to identify stellar masses larger than dynamical masses: \citet{Peralta2014} attribute an evolution in the stellar-to-dynamical mass being due to an evolving non-homology due to size evolution. 

One caveat with our analysis is that we have attributed all the evolution to an evolution of the luminous matter as opposed to the dark matter, and the discrepancies between $M_{dyn}$ and $M_{\star}$ mean that we are unable to estimate the dark matter fractions in these galaxies and so obtain a measure of how important this assumption might be. However, more detailed studies of individual galaxies have shown that the dark matter content of ETGs only dominates at large projected radii $R > R_e$ \citep[e.g.][]{Oldham2016a}, and so should not significantly affect stellar velocity dispersions that are measured in the central regions. It is therefore unlikely to be a significant problem in this study.

\subsection{Can we compare MACSJ0717 with Coma?}
The assumption at the foundation of this work is that the galaxies in MACSJ0717 represent an earlier evolutionary stage of the Coma galaxies: this allows us to compare their stellar populations and so make statements about the ages of their stars and the timescales of their formation. If this assumption is not valid, it could lead to systematic errors in our age calculations, so it is important to examine it closely. 

One possible problem could be the differing masses of the two clusters, with the X-ray luminosity of MACSJ0717 being more than three times greater than that of Coma -- compare $L_{MACS} = 24.6\times 10^{44} \text{ ergs}^{-1}$ \citep{ebeling07} with $L_{Coma} =  7.21 \times 10^{44} \text{ ergs}^{-1}$ \citep{Ebeling1996}. In the hierarchical paradigm, more massive dark matter haloes like that of MACSJ0717 are expected to collapse earlier and so have older stars. Further, if growth is accelerated in higher-mass, higher-density systems as we have suggested, this could also lead to inconsistencies in our framework. However, the evolution that we infer is sufficiently small that even an underlying age difference of $\sim 1$ Gyr would not make it significant. It is therefore extremely unlikely that either of these effects would bias our inference on scales larger than our uncertainties. We also note that the good agreement between our different age measurements suggests that the framework we have set up is consistent.

\section{Conclusions}
\label{sec:conclusions}
We have constructed the colour-magnitude relation, the fundamental plane and the $M_{dyn}/L-M_{dyn}$, $M_{\star}-\sigma$ and $M_{\star}-R_e$ relations for galaxies in the cluster MACSJ0717, at $z \sim 0.5$, using archived data from the CLASH and Gemini databases, and for Coma using existing datasets. By analysing these evolution between these relations, we have reached the following conclusions.

\begin{enumerate}
 \item The galaxies fall on an fundamental plane and an $M_{dyn}/L-M_{dyn}$ relation which are offset relative to those of Coma. The luminosity evolution implied by these offsets is $\Delta m \sim 0.6$ mag, corresponding to a star formation epoch of $\sim 10$ Gyr followed by passive fading.

 \item The galaxies fall on $M_{\star}-\sigma$ and $M_{\star}-R_e$ relations which are only marginally offset from those of Coma. The structural evolution implied by this is minimal, with $R_e(z) \sim (1+z)^{-0.40\pm0.32}$ and $\sigma(z) \sim (1+z)^{0.09 \pm 0.27}$, corresponding to galaxies which have undergone the majority of their evolution at earlier times.

 \item The fundamental plane and $M_{dyn}/L-M_{dyn}$, $M_{\star}-\sigma$ and $M_{\star}-R_e$ relations, modelled together, confirm these results. Importantly, the fact that all four relations can be modelled simulataneously and consistently implies that degeneracies between the physical parameters are not significant and that the physical scenario we have established, with evolution in luminosity, size and velocity dispersion, is consistent with and can fully account for the data. The fact that the inference from the independent colour-magnitude relation -- which is also based on different assumptions about the star formation histories of the galaxies -- is also consistent with these results further underlines this conclusion. 

 \item The small amount of structural evolution that we find in these galaxies is consistent with other studies of size evolution in cluster galaxies, but seems to be in tension with that found in studies of field ellipticals. This suggests that growth may be accelerated in high-density environments, where the rate of merging may be increased. If so, this is strong evidence that dry merging is a dominant channel of growth in these systems.

 \item Taken together, these results lead to a very clear picture in which these $z \sim 0.5$ galaxies have already experienced most of their star formation and structural evolution at earlier stages in their lives.

\end{enumerate}

\section*{Acknowledgements}
We thank the referee for their helpful comments. LJO thanks the Science and Technology Facilities Council (STFC) for the award of a studentship; RCWH was also supported by STFC [STFC grant numbers ST/H002456/1, ST/K00106X/1 \& ST/J002216/1]. RLD acknowledges travel and computer grants from Christ Church, Oxford and support from the Oxford Centre for Astrophysical Surveys which is funded by the Hintze Family Charitable Foundation.

This work was based on observations obtained at the Gemini Observatory (Program ID GN-2002B-Q-44; acquired through the Gemini Science Archive and processed using the Gemini IRAF package), which is operated by the Association of Universities for Research in Astronomy, Inc., under a cooperative agreement with the NSF on behalf of the Gemini partnership: the National Science Foundation (United States), the National Research Council (Canada), CONICYT (Chile), Ministerio de Ciencia, Tecnología e Innovación Productiva (Argentina), and Ministério da Ciência, Tecnologia e Inovação (Brazil).

This work was also based on observations made with the NASA/ESA Hubble Space Telescope, obtained from the Data Archive at the Space Telescope Science Institute, which is operated by the Association of Universities for Research in Astronomy, Inc., under NASA contract NAS 5-26555. These observations are associated with the CLASH Multi-Cycle Treasury Program (GO-12065).

\onecolumn
\section{Appendix}
\setcounter{table}{0}
\renewcommand{\thetable}{A\arabic{table}}

\setcounter{figure}{0} 
\renewcommand{\thefigure}{A\arabic{figure}}

\noindent\begin{minipage}{\textwidth}
    \centering
\includegraphics[trim=20 20 20 20,clip,width=\textwidth]{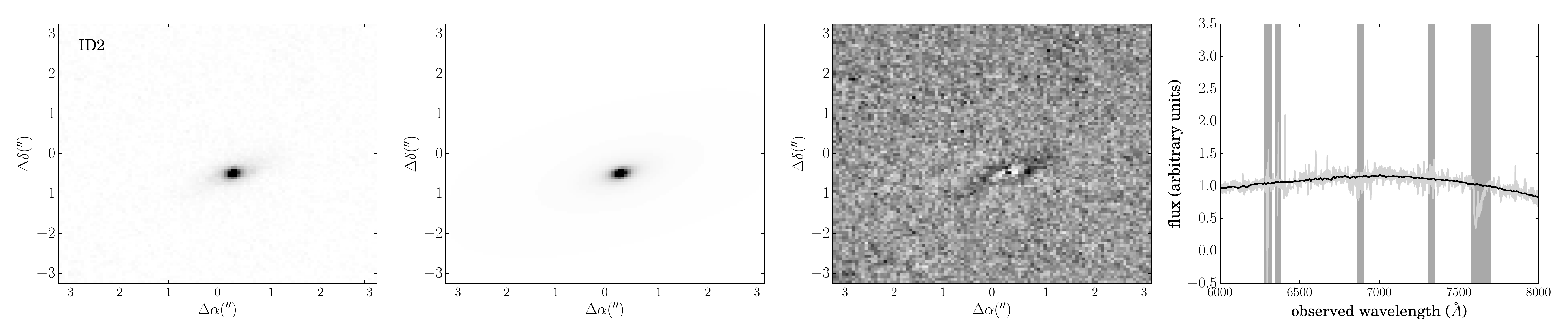}\hfill
\includegraphics[trim=20 20 20 20,clip,width=\textwidth]{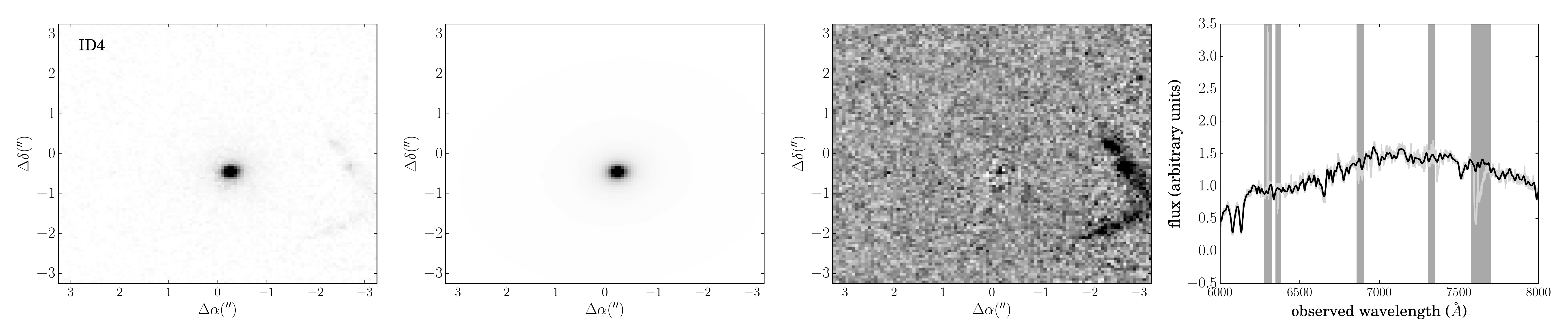}\hfill
\includegraphics[trim=20 20 20 20,clip,width=\textwidth]{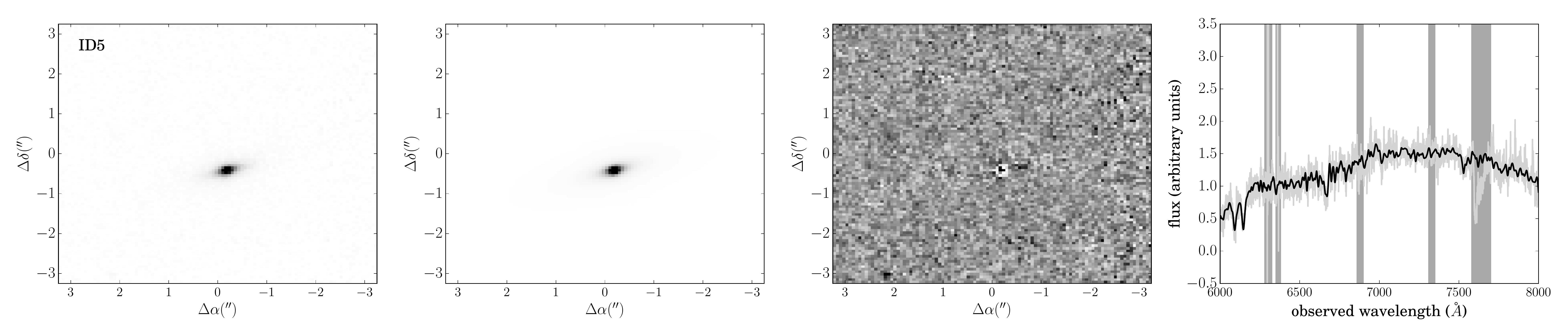}\hfill
\includegraphics[trim=20 20 20 20,clip,width=\textwidth]{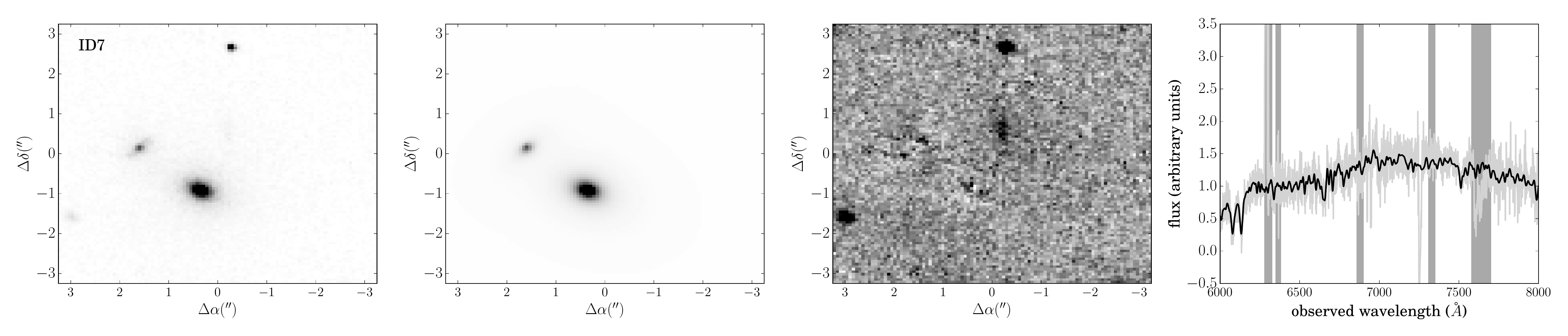}\hfill
\includegraphics[trim=20 20 20 20,clip,width=\textwidth]{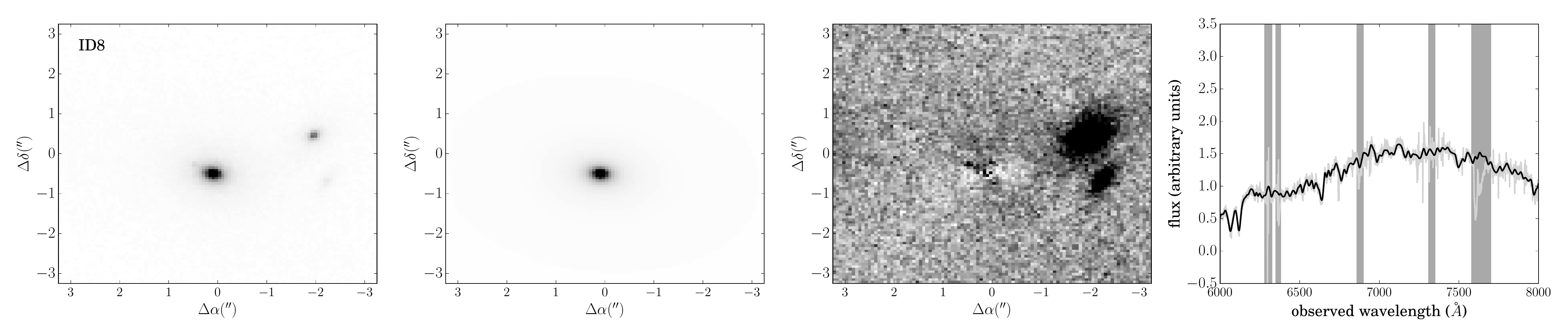}\hfill
\includegraphics[trim=20 20 20 20,clip,width=\textwidth]{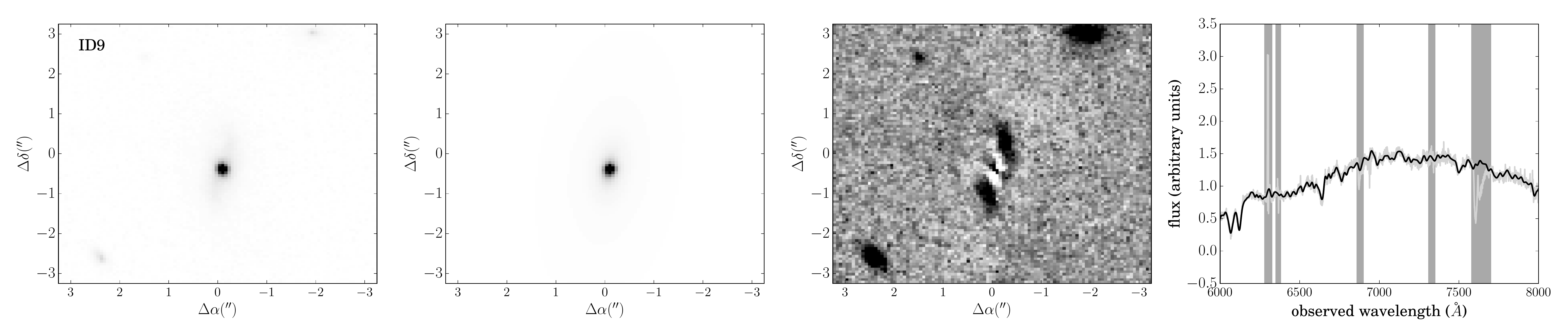}\hfill

\end{minipage}
\begin{figure}
 \centering
\subfigure{\includegraphics[trim=20 20 20 20,clip,width=\textwidth]{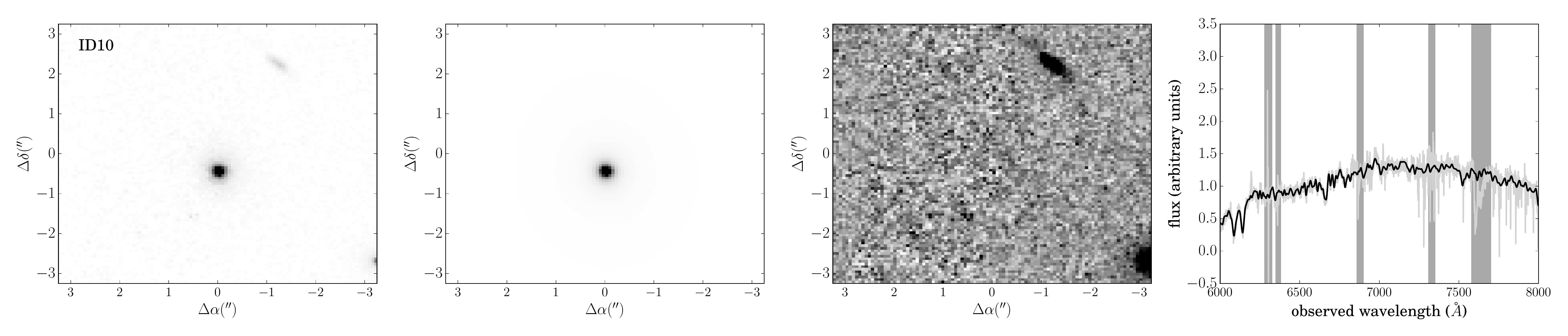}}\hfill
\subfigure{\includegraphics[trim=20 20 20 20,clip,width=\textwidth]{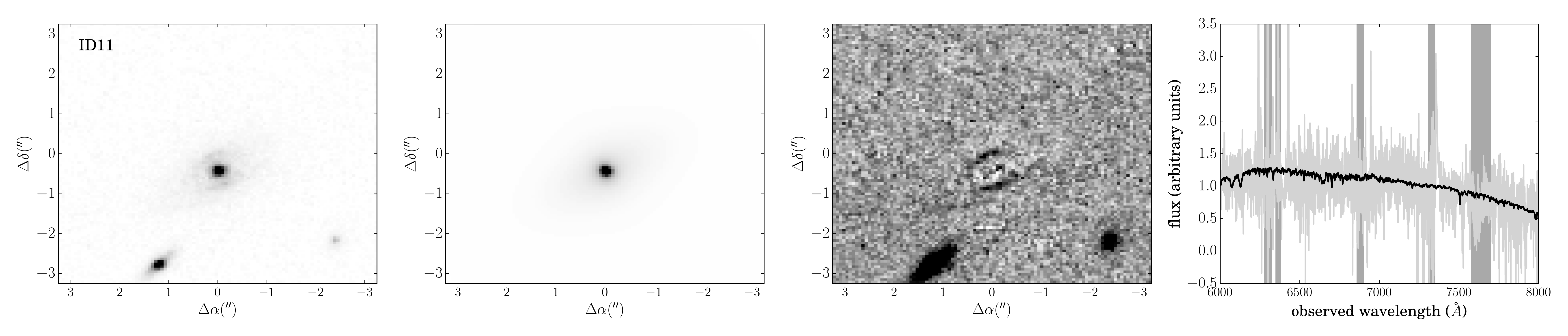}}\hfill
\subfigure{\includegraphics[trim=20 20 20 20,clip,width=\textwidth]{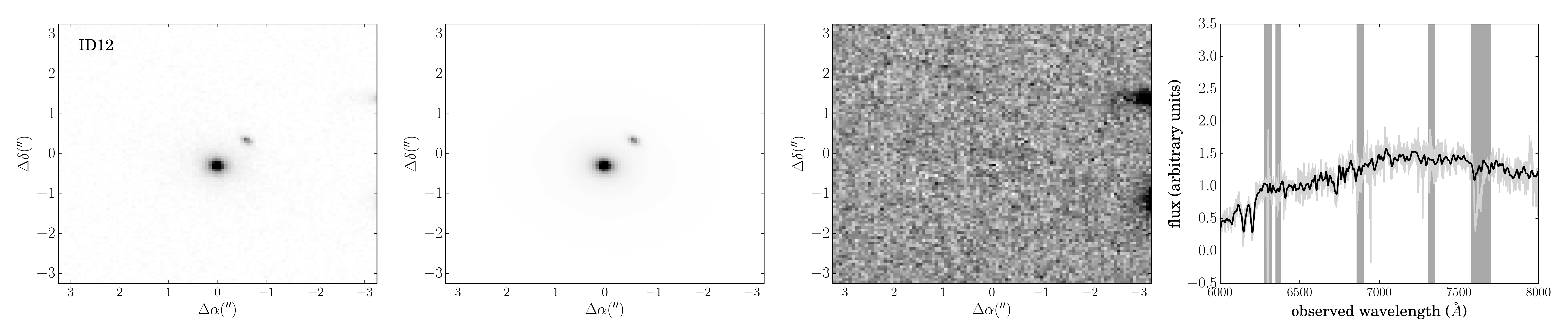}}\hfill
\subfigure{\includegraphics[trim=20 20 20 20,clip,width=\textwidth]{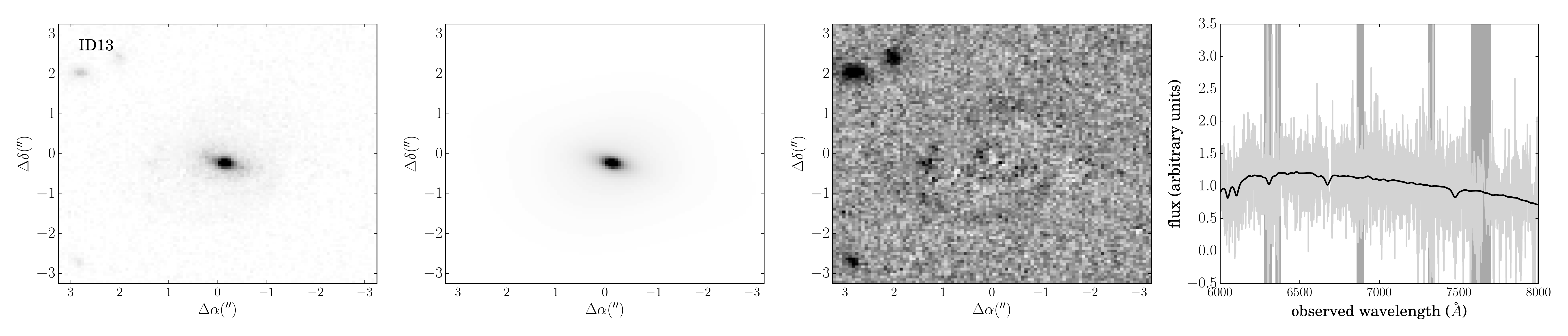}}\hfill
\subfigure{\includegraphics[trim=20 20 20 20,clip,width=\textwidth]{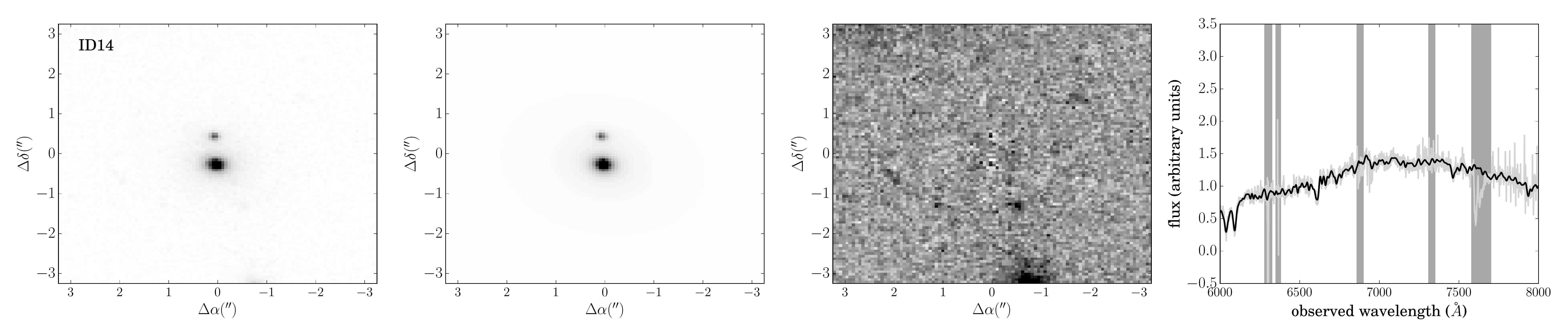}}\hfill
\subfigure{\includegraphics[trim=20 20 20 20,clip,width=\textwidth]{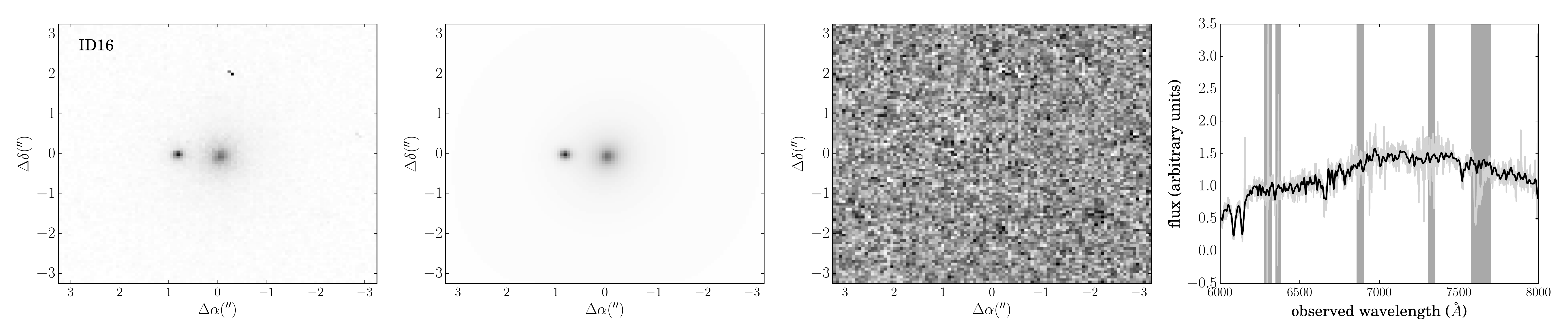}}\hfill
\end{figure}

\begin{figure}
 \centering
\subfigure{\includegraphics[trim=20 20 20 20,clip,width=\textwidth]{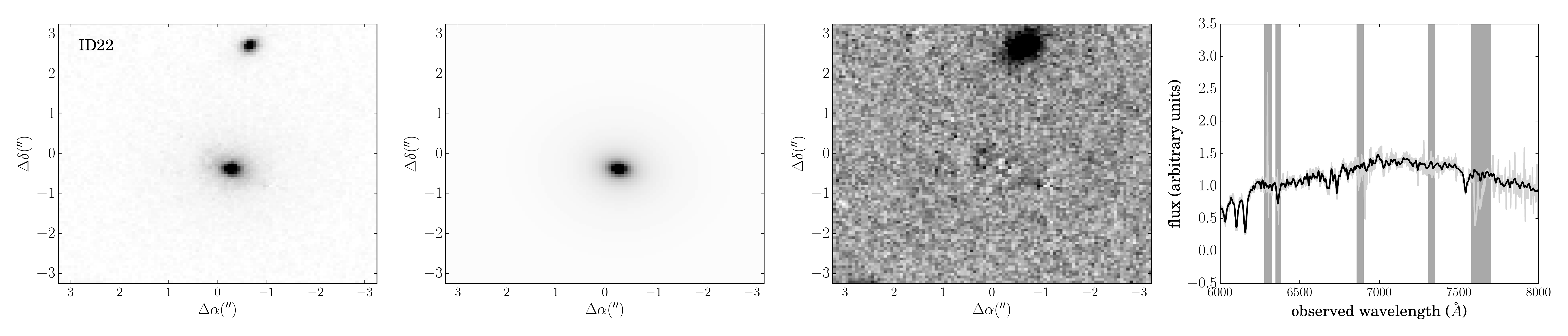}}\hfill
\subfigure{\includegraphics[trim=20 20 20 20,clip,width=\textwidth]{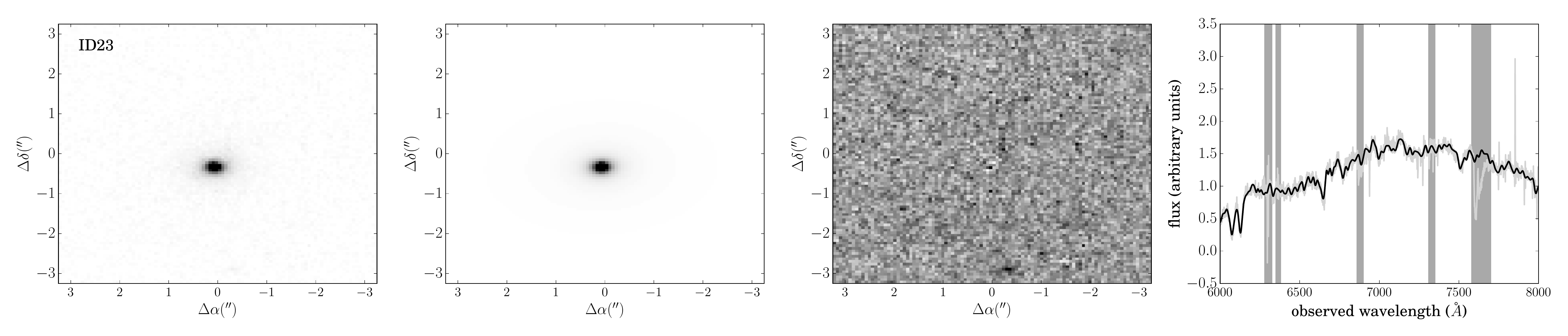}}\hfill
\subfigure{\includegraphics[trim=20 20 20 20,clip,width=\textwidth]{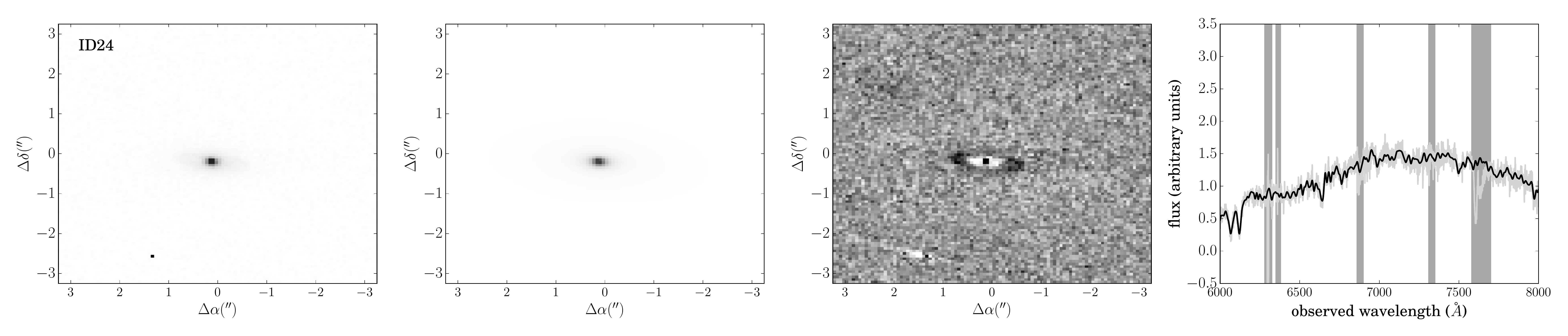}}\hfill
\subfigure{\includegraphics[trim=20 20 20 20,clip,width=\textwidth]{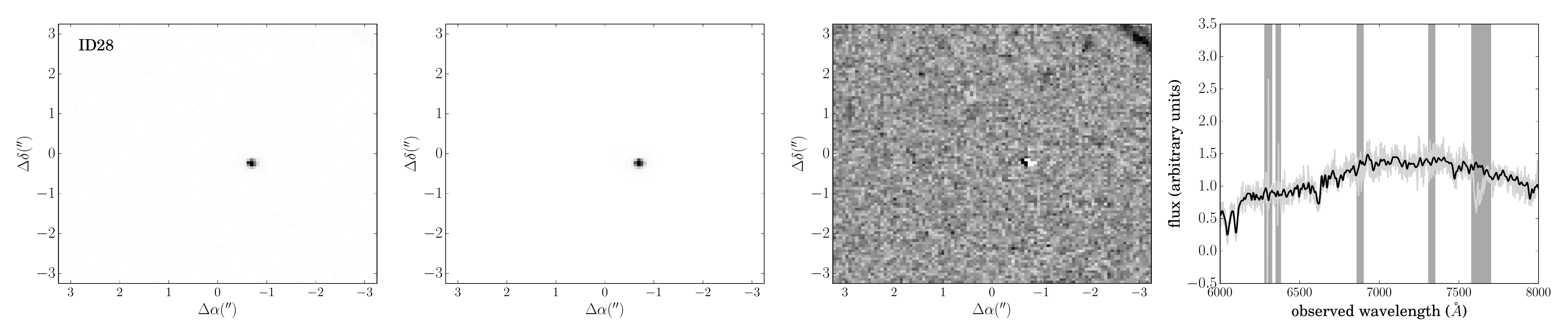}}\hfill
\label{fig:sersices}
\caption{For each object in our final MACSJ0717 sample, from left to right: HST/ACS F625W data; model; signal-to-noise residuals; fitted spectrum. Image cutouts are 6.5 $''$ on a side; spectra show the extracted spectrum in grey and the model in black, with masked regions in dark grey.}
\end{figure}

\begin{table}
 \centering
\begin{tabular}{ccc}\hline
 ID & $U$ / mag & $B$ / mag \\\hline
2 & $-19.65 \pm 0.08$ & $-21.09 \pm 0.08$ \\
4 & $-19.88 \pm 0.09$ & $-21.32 \pm 0.09$ \\
5 & $-19.89 \pm 0.05$ & $-21.25 \pm 0.05$ \\
7 & $-20.61 \pm 0.06$ & $-22.07 \pm 0.06$ \\
8 & $-22.62 \pm 0.08$ & $-24.08 \pm 0.08$ \\
9 & $-21.68 \pm 0.09$ & $-22.94 \pm 0.09$ \\
10 & $-20.20 \pm 0.06$ & $-21.30 \pm 0.06$ \\ 
11 & $-20.59 \pm 0.06$ & $-21.92 \pm 0.06$ \\
12 & $-20.39 \pm 0.08 $ & $-21.49 \pm 0.08$\\
13 & $-20.39 \pm 0.08 $ & $-21.64\pm 0.08$ \\
14 & $-21.06 \pm 0.10$ & $-22.17 \pm 0.10$ \\
16 & $-21.59 \pm 0.06$ & $-23.05 \pm 0.06$ \\
22 & $-21.04 \pm 0.09$ & $-22.15 \pm 0.09$ \\
23 & $-20.01 \pm 0.06$ & $-21.12 \pm 0.06$ \\
24 & $-19.96 \pm 0.09$ & $-21.32 \pm 0.09$ \\
28 & $-19.42 \pm 0.18$ & $-19.94 \pm 0.18$ \\\hline
\end{tabular}
\caption{Rest-frame absolute magnitudes for the MACSJ0717 galaxies in the Johnson $U$ and $B$ bands. We calculate these by fitting \citet{bruzual} SPS models to the F625W and F475W magnitudes and colours, assuming a Salpeter IMF and solar metallicity as in Section 4; this allows us to infer a stellar mass and age. We then evaluate these models in the rest-frame $U$ and $V$ filters.}
\end{table}

\twocolumn


\begin{thebibliography}{}

\bibitem[Auger et al.(2010)]{Auger2010} Auger, M.~W., Treu, T., 
Gavazzi, R., et al.\ 2010, \apjl, 721, L163 

\bibitem[\protect\citeauthoryear{Baum}{Baum}{1959}]{baum}
Baum W.~A.,  1959, \pasp, 71, 106

\bibitem[\protect\citeauthoryear{Bender, Ziegler \& Bruzual}{Bender
  et~al.}{1996}]{bender98}
Bender R.,  Ziegler B.,    Bruzual G.,  1996, \apj, 463

\bibitem[\protect\citeauthoryear{Bertin \& Arnouts}{Bertin}{1996}]{Bertin1996} 
Bertin, E., \& Arnouts, S.\ 1996, \aaps, 117, 393 


\bibitem[\protect\citeauthoryear{Bouwens et al.}{Bouwens et al.}{2004}]{Bouwens2004}
Bouwens, R.~J., Illingworth, G.~D., Blakeslee, J.~P., Broadhurst, T.~J., 
\& Franx, M.\ 2004, \apjl, 611, L1 


\bibitem[\protect\citeauthoryear{Bower, Kodama \& Terlevich}{Bower
  et~al.}{1998}]{bower98}
Bower R.~G.,  Kodama T.,    Terlevich A.,  1998, \mnras, 299, 1193

\bibitem[\protect\citeauthoryear{Bower, Lucey \& Ellis}{Bower
  et~al.}{1992}]{bower92}
Bower R.~G.,  Lucey J.~R.,    Ellis R.~S.,  1992, \mnras, 254, 601

\bibitem[\protect\citeauthoryear{Bruzual \& Charlot}{Bruzual \&
  Charlot}{2003}]{bruzual}
Bruzual G.,  Charlot S.,  2003, \mnras, 344, 1000

Butcher, H., \& Oemler, A., Jr.\ 1984, \apj, 285, 426 

\bibitem[Cappellari \& Emsellem(2004)]{cappellari04} Cappellari, M., \& Emsellem, E.\ 2004, \pasp, 116, 138 


\bibitem[\protect\citeauthoryear{Cappellari et al.}{Cappellari et~al.}{2006}]{cappellari06}
Cappellari M., et al. 2006, \mnras, 366, 1126


\bibitem[\protect\citeauthoryear{Cenarro \& Trujillo}{Cenarro \&
  Trujillo}{2009}]{Cenarro2009}
Cenarro A.~J.,  Trujillo I.,  2009, p.~5

\bibitem[Chabrier(2003)]{Chabrier} Chabrier, G.\ 2003, \pasp, 115, 763 


\bibitem[\protect\citeauthoryear{Clemens, Bressan, Panuzzo, Rampazzo, Silva,
  Buson \& Granato}{Clemens et~al.}{2009}]{clemens09}
Clemens M.~S.,  Bressan A.,  Panuzzo P.,  Rampazzo R.,  Silva L.,  Buson L.,
  Granato G.~L.,  2009, \mnras, 392, 982

\bibitem[Delaye et al.(2014)]{Delaye2014} Delaye, L., Huertas-Company, M., Mei, S., et al.\ 2014, \mnras, 441, 203 


\bibitem[\protect\citeauthoryear{Djorgovski \& Davis}{Djorgovski \&
  Davis}{1987}]{djorgovski}
Djorgovski S.,  Davis M.,  1987, \apj, 313, 59

\bibitem[\protect\citeauthoryear{Dressler, Lynden-Bell, Burstein, Davies,
  Faber, Terlevich \& Wegner}{Dressler et~al.}{1987}]{dressler87}
Dressler A.,  Lynden-Bell D.,  Burstein D.,  Davies R.~L.,  Faber S.~M.,
  Terlevich R.,    Wegner G.,  1987, \apj, 313, 42

\bibitem[\protect\citeauthoryear{Ebeling et al.}{Ebeling et al.}{1996}]{Ebeling1996}
Ebeling, H., Voges, W., Bohringer, H., et al.\ 1996, \mnras, 281, 799 

\bibitem[\protect\citeauthoryear{Ebeling, Barrett, Donovan, Ma, Edge \& van
  Speybroeck}{Ebeling et~al.}{2007}]{ebeling07}
Ebeling H.,  Barrett E.,  Donovan D.,  Ma C.-J.,  Edge A.~C.,    van Speybroeck
  L.,  2007, \apj, 661

\bibitem[\protect\citeauthoryear{Ellis, Smail, Dressler, Couch, Oemler~Jr.,
  Butcher \& Sharples}{Ellis et~al.}{1997}]{ellis97}
Ellis R.~S.,  Smail I.,  Dressler A.,  Couch W.~J.,  Oemler~Jr. A.,  Butcher
  H.,    Sharples R.~M.,  1997, \apj, 483, 582

\bibitem[Fan et al.(2010)]{Fan2010} Fan, L., Lapi, A., Bressan, 
A., et al.\ 2010, \apj, 718, 1460 

\bibitem[\protect\citeauthoryear{Foreman-Mackey, Hogg, Lang \&
  Goodman}{Foreman-Mackey et~al.}{2013}]{ForemanMackey2013}
Foreman-Mackey D.,  Hogg D.~W.,  Lang D.,    Goodman J.,  2013, Publ. Astron.
  Soc. Pacific, 125, 306

\bibitem[\protect\citeauthoryear{Frei \& Gunn}{Frei \& Gunn}{1994}]{FreiGunn1994}
Frei, Z., \& Gunn, J.~E.\ 1994, \aj, 108, 1476 

\bibitem[\protect\citeauthoryear{Hanuschik}{Hanuschik}{2003}]{Hanuschik2003}
Hanuschik, R.~W.\ 2003, \aap, 407, 1157 

\bibitem[Hogg et al.(2002)]{Hogg2002} Hogg, D.~W., Baldry, I.~K., Blanton, M.~R., \& Eisenstein, D.~J.\ 2002, arXiv:astro-ph/0210394 


\bibitem[\protect\citeauthoryear{Hogg, Bovy \& Lang}{Hogg
  et~al.}{2010}]{hogg10}
Hogg D.~W.,  Bovy J.,    Lang D.,  2010, ArXiv e--prints

\bibitem[\protect\citeauthoryear{Holden, van~der Wel, Kelson, Franx \&
  Illingworth}{Holden et~al.}{2010}]{holden10}
Holden B.~P.,  van~der Wel A.,  Kelson D.~D.,  Franx M.,    Illingworth G.~D.,
  2010, \apj, 724, 714


\bibitem[\protect\citeauthoryear{Houghton, Davies, Dalla~Bontà \&
  Masters}{Houghton et~al.}{2012}]{houghton}
Houghton R. C.~W.,  Davies R.~L.,  Dalla~Bontà E.,    Masters R.,  2012,
  \mnras, 423, 256

\bibitem[\protect\citeauthoryear{{J{\o}rgensen}, {Franx} \&
  {Kjaergaard}}{{J{\o}rgensen} et~al.}{1995}]{jorgensen95}
{J{\o}rgensen} I.,  {Franx} M.,    {Kjaergaard} P.,  1995, \mnras, 273, 1097


\bibitem[\protect\citeauthoryear{J{\o}rgensen, Franx, Hjorth \& van
  Dokkum}{J{\o}rgensen et~al.}{1999a}]{jorgensen99}
J{\o}rgensen I.,  Franx M.,  Hjorth J.,    van Dokkum P.~G.,  1999a, \mnras, 308,
  833

\bibitem[\protect\citeauthoryear{J{\o}rgensen, Chiboucas, Flint, Bergmann, Barr \&
  Davies}{J{\o}rgensen et~al.}{2006}]{jorgensen06}
J{\o}rgensen I.,  Chiboucas K.,  Flint K.,  Bergmann M.,  Barr J.,    Davies R.,
  2006, \apj, 639

\bibitem[\protect\citeauthoryear{J{\o}rgensen \& Chiboucas}{J{\o}rgensen \& Chiboucas}{2013}]{JorgensenChiboucas2013}
J{\o}rgensen, I., \& Chiboucas, K.\ 2013, \aj, 145, 77 

\bibitem[\protect\citeauthoryear{J{\o}rgensen et al.}{J{\o}rgensen et al.}{2014}]{Jorgensen2014}
J{\o}rgensen, I., Chiboucas, K., Toft, S., et al.\ 2014, \aj, 148, 117 

\bibitem[\protect\citeauthoryear{Kelly}{Kelly}{2007}]{Kelly2007}
Kelly B.~C.,  2007, \apj, 665, 1489

\bibitem[\protect\citeauthoryear{Kodama \& Arimoto}{Kodama \&
  Arimoto}{1997}]{kodama}
Kodama T.,  Arimoto N.,  1997, \aa, 320, 41

\bibitem[Kroupa(2001)]{Kroupa} Kroupa, P.\ 2001, \mnras, 322, 231 


\bibitem[Lani et al.(2013)]{Lani2013} Lani, C., Almaini, O., Hartley, W.~G., et al.\ 2013, \mnras, 435, 207 


\bibitem[\protect\citeauthoryear{Mehlert, Thomas, Saglia \& Bender R.}{Mehlert
  et~al.}{2003}]{mehlert03}
Mehlert D.,  Thomas D.,  Saglia R.~P.,    Bender R. W.~G.,  2003, A\& A, 407,
  423

\bibitem[\protect\citeauthoryear{Mei, Holden, Blakeslee, Ford, Franx, Homeier,
  Illingworth, Jee, Overzier, Postman, Rosati, Van~der Wel \& Bartlett}{Mei
  et~al.}{2009}]{mei}
Mei S.,  Holden B.~P.,  Blakeslee J.,  Ford H.~C.,  Franx M.,  Homeier N.~L.,
  Illingworth G.~D.,  Jee M.~J.,  Overzier R.,  Postman M.,  Rosati P.,
  Van~der Wel A.,    Bartlett J.~G.,  2009, \apj, 690, 42


\bibitem[\protect\citeauthoryear{Naab et al.}{Naab et al.}{2009}]{Naab2009} 
Naab, T., Johansson, P.~H., \& Ostriker, J.~P.\ 2009, \apjl, 699, L178 

\bibitem[Newman et al.(2014)]{Newman2014} Newman, A.~B., Ellis, R.~S., Andreon, S., et al.\ 2014, \apj, 788, 51 


\bibitem[\protect\citeauthoryear{Oldham \& Auger}{Oldham \& Auger}{2016a}]{Oldham2016a}
Oldham, L.~J., \& Auger, M.~W.\ 2016, \mnras, 457, 421

\bibitem[\protect\citeauthoryear{Oldham et al.}{Oldham et al.}{2016b}]{Oldham2016b}
Oldham, L.~J., et al. \ 2016, \mnras, submitted

\bibitem[Peralta de Arriba et al.(2014)]{Peralta2014} Peralta de Arriba, L., Balcells, M., Falc{\'o}n-Barroso, J., \& Trujillo, I.\ 2014, \mnras, 440, 1634 


\bibitem[Postman et al.(2012)]{Postman2012} Postman, M., Coe, D., 
Ben{\'{\i}}tez, N., et al.\ 2012, \apjs, 199, 25 

\bibitem[\protect\citeauthoryear{Saglia, Sanchez-Blazquez, Bender, Simard,
  Desai, Aragon-Salamanca, Milvang-Jensen \& Halliday}{Saglia
  et~al.}{2010}]{saglia10}
Saglia R.~P.,  Sanchez-Blazquez P.,  Bender R.,  Simard L.,  Desai V.,
  Aragon-Salamanca A.,  Milvang-Jensen B.,    Halliday C. e.~a.,  2010, A\& A,
  524

\bibitem[\protect\citeauthoryear{Sandage}{Sandage}{1972}]{sandage}
Sandage A.,  1972, \apj, 176, 21

\bibitem[\protect\citeauthoryear{Saracco et al.}{Saracco et al.}{2014}]{Saracco2014} 
Saracco, P., Casati, A., Gargiulo, A., et al.\ 2014, \aap, 567, A94 

\bibitem[Shen et al.(2003)]{Shen2003} Shen, S., Mo, H.~J., 
White, S.~D.~M., et al.\ 2003, \mnras, 343, 978 


\bibitem[\protect\citeauthoryear{Stanford, Eisenhardt \& Dickinson}{Stanford
  et~al.}{1995}]{stanford95}
Stanford S.~A.,  Eisenhardt P.~R.,    Dickinson M.,  1995, \apj, 450, 512

\bibitem[\protect\citeauthoryear{Stanford, Eisenhardt \& Dickinson}{Stanford
  et~al.}{1998}]{stanford98}
Stanford S.~A.,  Eisenhardt P.~R.,    Dickinson M.,  1998, \apj, 492, 461

\bibitem[Stockton et al.(2014)]{Stockton2014} Stockton, A., Shih, 
H.-Y., Larson, K., \& Mann, A.~W.\ 2014, \apj, 780, 134 


\bibitem[\protect\citeauthoryear{Stott et al.}{Stott et al.}{2011}]{Stott2011}
Stott, J.~P., Collins, C.~A., Burke, C., Hamilton-Morris, V., \& Smith, G.~P.\ 2011, \mnras, 414, 445 


\bibitem[\protect\citeauthoryear{Terlevich, Caldwell \& Bower}{Terlevich
  et~al.}{2001}]{terlevich01}
Terlevich A.~I.,  Caldwell N.,    Bower R.~G.,  2001, \mnras, 326, 1547


\bibitem[Tody(1993)]{Tody1993} Tody, D.\ 1993, Astronomical Data Analysis Software and Systems II, 52, 173 


\bibitem[\protect\citeauthoryear{Trujillo, Conselice, Bundy, Cooper, Eisenhardt
  \& Ellis}{Trujillo et~al.}{2007}]{Trujillo2007}
Trujillo I.,  Conselice C.~J.,  Bundy K.,  Cooper M.~C.,  Eisenhardt P.,
  Ellis R.~S.,  2007, \mnras, 382, 109

\bibitem[\protect\citeauthoryear{Valdes, Gupta, Rose, Singh \& Bell}{Valdes
  et~al.}{2004}]{valdes}
Valdes F.,  Gupta R.,  Rose J.~A.,  Singh H.~P.,    Bell D.~J.,  2004, \apjs,
  152, 251

\bibitem[Valentinuzzi et al.(2010)]{Valentinuzzi2010} Valentinuzzi, T., 
Poggianti, B.~M., Saglia, R.~P., et al.\ 2010, \apjl, 721, L19 


\bibitem[\protect\citeauthoryear{van~der Wel, Holden, Zirm, Franx, Rettura,
  Illingworth \& Ford}{van~der Wel et~al.}{2008}]{VanderWel2008}
van~der Wel A.,  Holden B.~P.,  Zirm A.~W.,  Franx M.,  Rettura A.,
  Illingworth G.~D.,    Ford H.~C.,  2008, \apj, 688, 48

\bibitem[van der Wel et al.(2014)]{vanderWel2014} van der Wel, A., Franx, M., van Dokkum, P.~G., et al.\ 2014, \apj, 788, 28 


\bibitem[\protect\citeauthoryear{van Dokkum \& Franx}{van Dokkum \&
  Franx}{1996}]{vandokkum96}
van Dokkum P.~G.,  Franx M.,  1996, \mnras, 281, 985

\bibitem[\protect\citeauthoryear{van Dokkum \& Ellis}{van Dokkum \&
  Ellis}{2003}]{vanDokkum2003}
van Dokkum P.~G.,  Ellis R.~S.,  2003, \apj, 592


\bibitem[van Dokkum et al.(2008)]{vanDokkum2008} van Dokkum, P.~G., 
Franx, M., Kriek, M., et al.\ 2008, \apjl, 677, L5 


\bibitem[\protect\citeauthoryear{van Dokkum, Kriek \& Franx}{van Dokkum
  et~al.}{2009}]{vanDokkum2009}
van Dokkum P.~G.,  Kriek M.,    Franx M.,  2009


\bibitem[\protect\citeauthoryear{Visvanathan \& Sandage}{Visvanathan \&
  Sandage}{1977}]{visvanathan}
Visvanathan N.,  Sandage A.,  1977, \apj, 216, 214

\bibitem[Wolf et al.(2010)]{Wolf2010} Wolf, J., Martinez, G.~D., Bullock, J.~S., et al.\ 2010, \mnras, 406, 1220 

\end{thebibliography}
\end{document}